\documentclass[intlimits,twoside,a4paper]{article}

\usepackage[cp1251]{inputenc}

\usepackage[eqsecnum]{cmpj3}

\usepackage{bm}

\issue{2026}{29}{2}{23502}
\doinumber{10.5488/CMP.29.23502}
\title[An elastic model of confined hydrogel particles]%
{An elastic model of confined hydrogel particles with competing entropic and energetic networks\thanks{This work is dedicated to the memory of Stefan Soko\l{}owski, whose insightful molecular-dynamics studies of granular matter continue to inspire research on collective behavior in soft and particulate systems.}}
\author[A. Huerta, L. A. P\'erez, A. Trokhymchuk]{A. Huerta\orcid{0000-0003-2435-6397}\refaddr{label1}\thanks{On sabbatical leave at Instituto de F\'isica, Universidad Nacional Aut\'onoma de M\'exico, C.P. 04510 Ciudad de M\'exico, M\'exico},
        L. A. P\'erez\orcid{0000-0003-0983-8486}\refaddr{label2},
        A. Trokhymchuk\orcid{0000-0003-2257-0779}\refaddr{label3}\thanks{Corresponding author: \email{adt@icmp.lviv.ua}.}}
\addresses{
\addr{label1} Facultad de F\'isica, Universidad Veracruzana, Campus Arco Sur, Paseo 112, C. P. 91097 Xalapa, M\'exico
\addr{label2} Instituto de F\'isica, Universidad Nacional Aut\'onoma de M\'exico, C.P. 04510 Ciudad de M\'exico, M\'exico
\addr{label3} Yukhnovskii Institute for Condensed Matter Physics of the National Academy of Sciences of Ukraine, 1~Svientsitskii Str., 79011 Lviv, Ukraine
}

\date{Received 3 February 2026; revised 7 March 2026; accepted 11 March 2026; published 29 June 2026}

\begin{document}

\maketitle

\begin{abstract}
This work presents an elastic model to study the interplay between entropic and energetic networks in confined hydrogel particles. We consider a quasi-two-dimensional system composed of spherical hydrogel beads confined in a circular container, where particle growth occurs through hydration. Based on experimental observations, an elastic potential is introduced to model interactions between particles and between particles and the confining wall. 
Computational simulations based on energy minimization identify the lowest-energy configurations adopted during growth. Analysis of the resulting energy landscapes reveals emergent self-organization, adaptability, and cooperativity arising from the competition between entropic and energetic networks.
\printkeywords
%
\end{abstract}

\section{Introduction}

\begin{figure*}[!ht]
\includegraphics[width=0.495\textwidth]{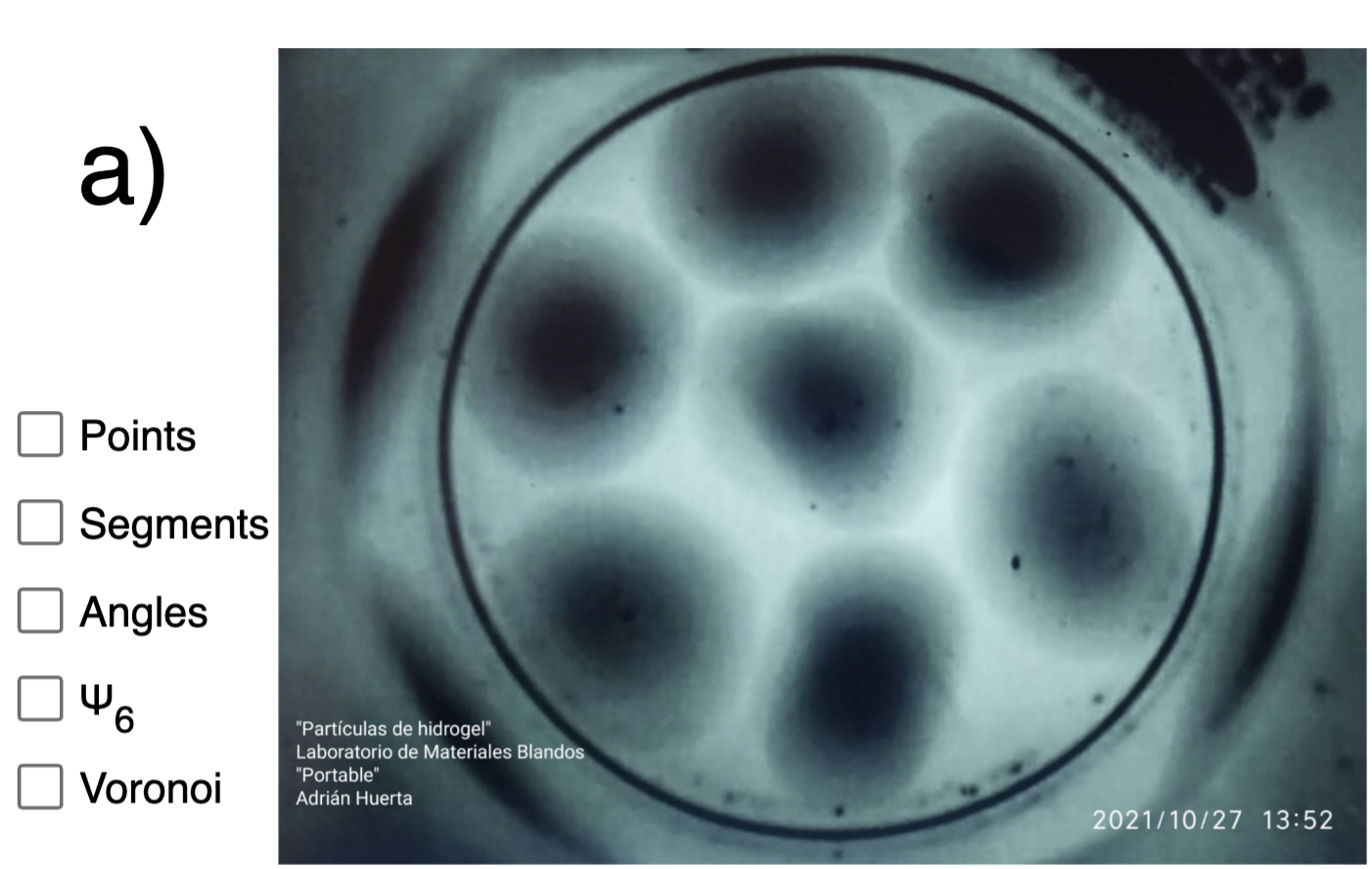}
\includegraphics[width=0.495\textwidth]{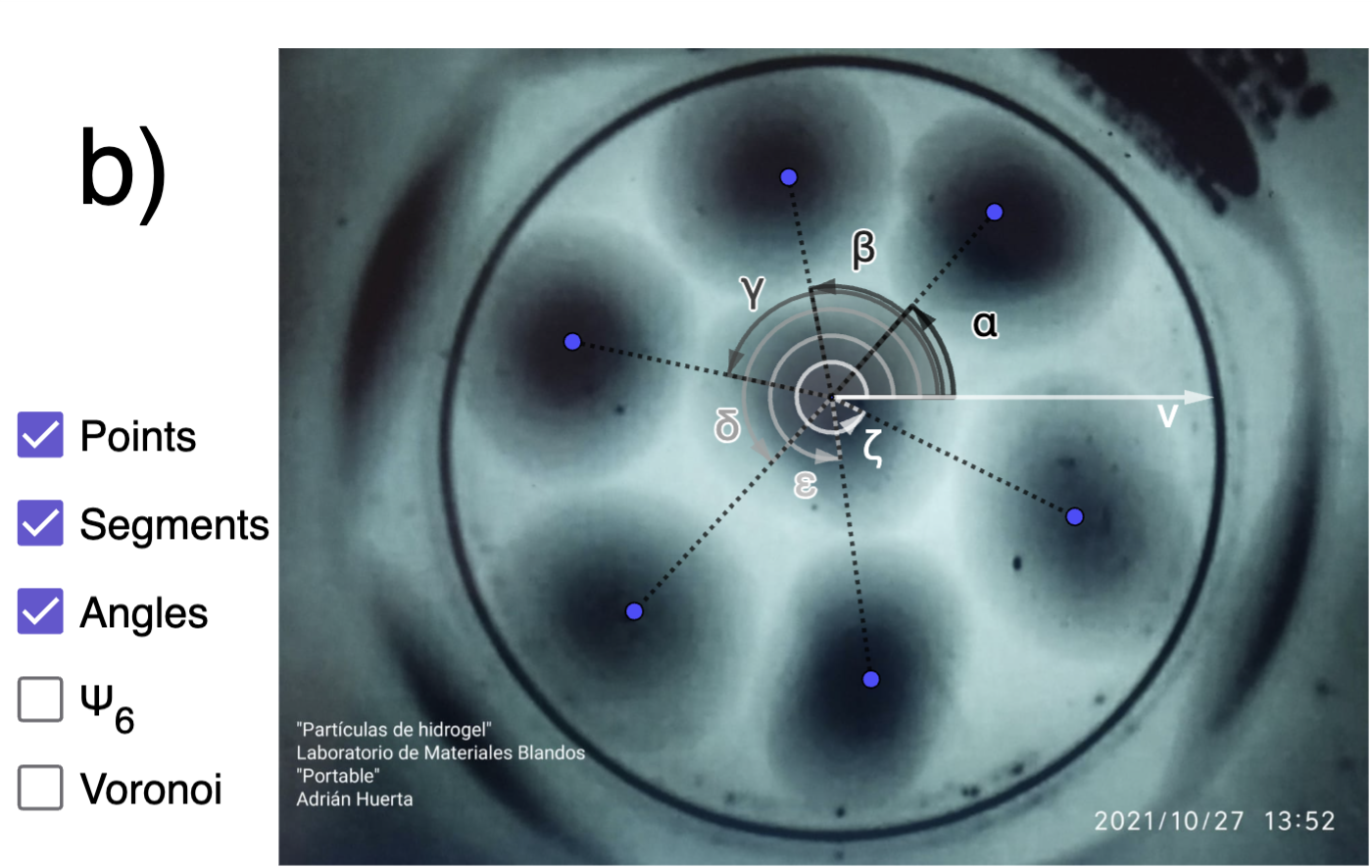}
\includegraphics[width=0.495\textwidth]{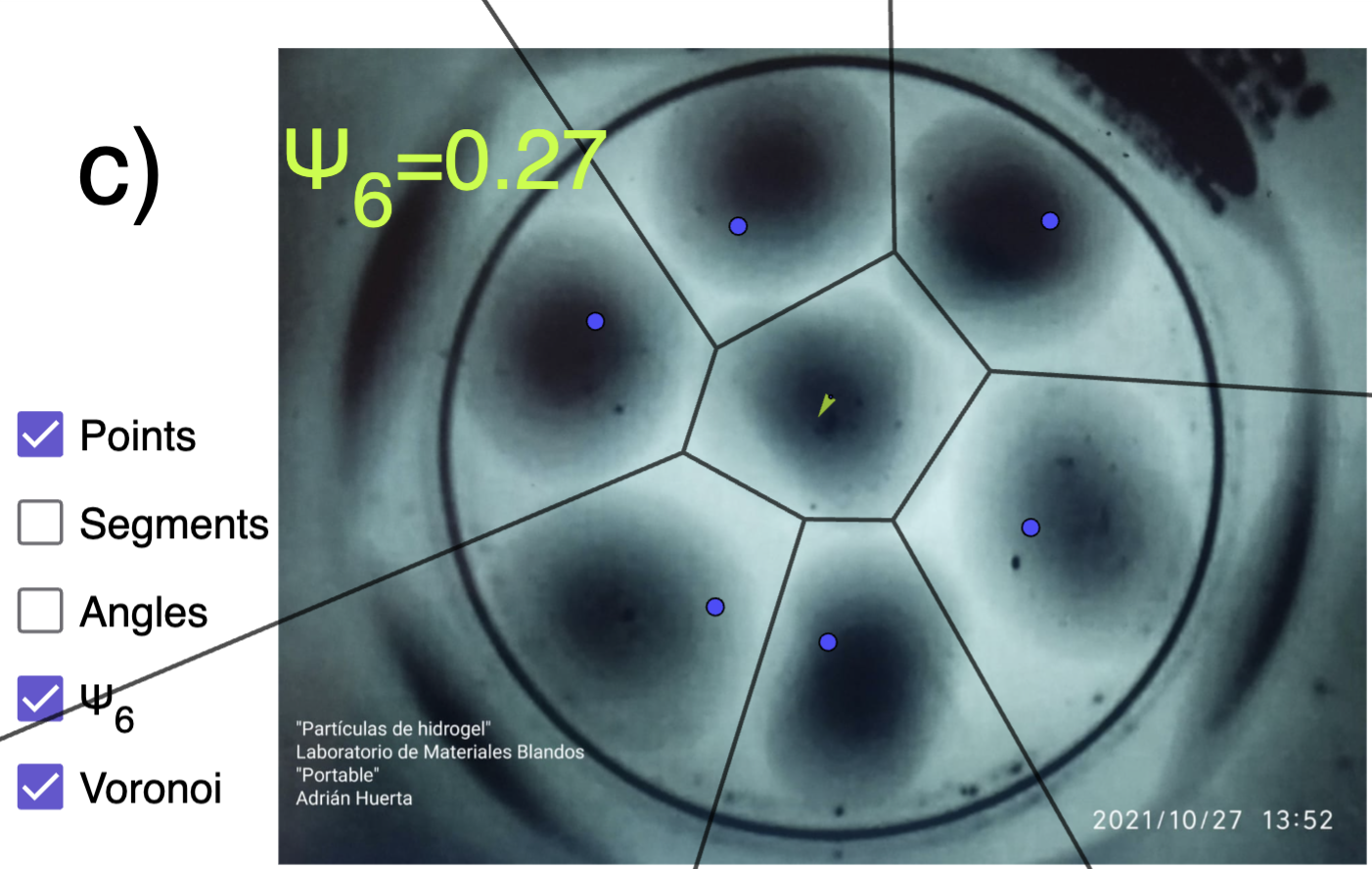}
\includegraphics[width=0.495\textwidth]{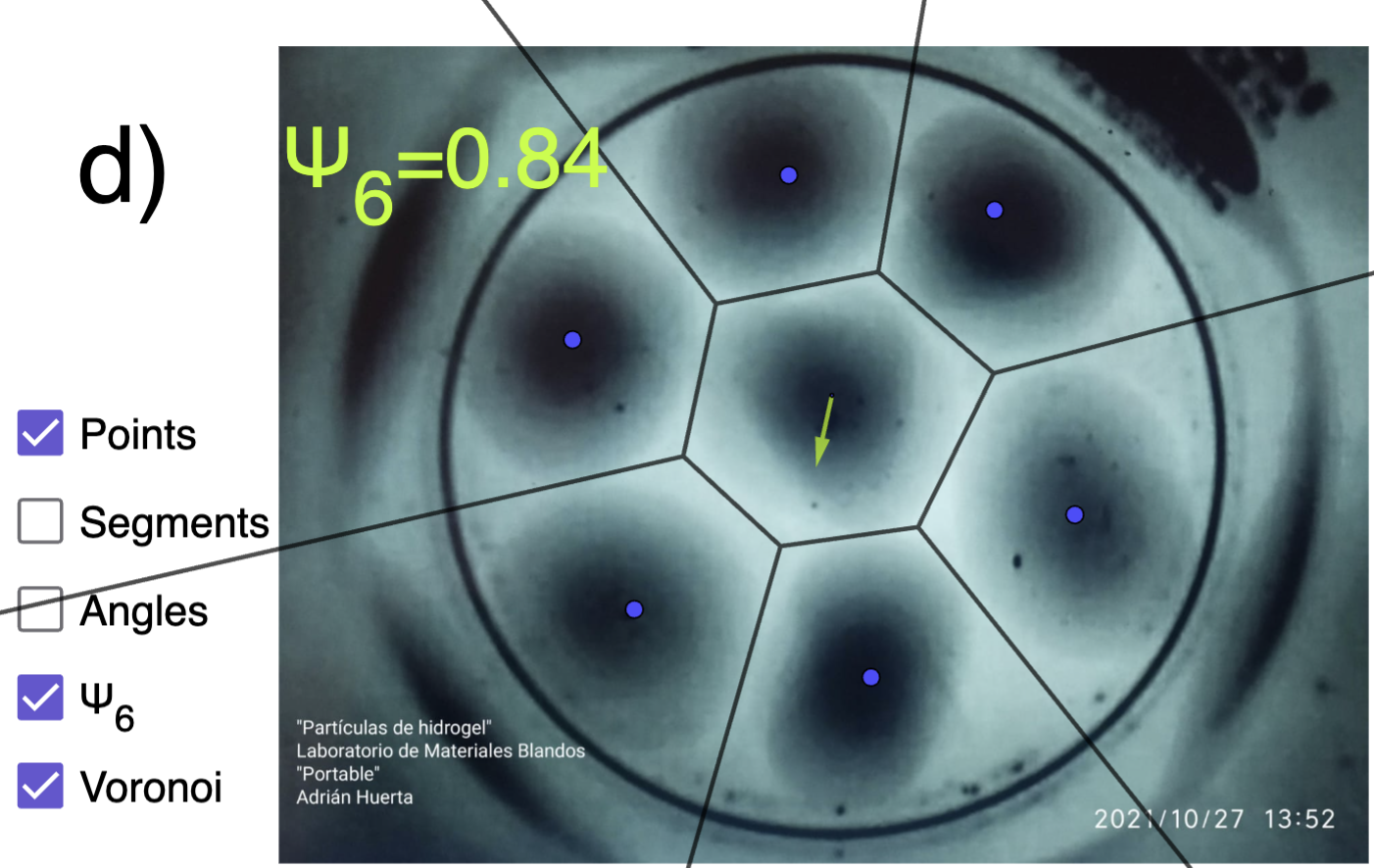}
 \caption{(Colour online) a) Structure formed by seven hydrogel particles growing in water,
 which grew confined within a glass, completely submerged. b) The approximate centers of the particles are indicated by points, as well as the segments connecting them, which will help us calculate the angles formed with the $x$-axis, necessary to evaluate the order parameter $\Psi_6$; c) Voronoi tessellation used to identify the domains and neighborhood of the particles; d) The points corresponding to the centers of the seven hydrogel particles are adjusted so that the lines of the polygons coincide with the deformations of the particles. Its operation can be seen at the following link as supplementary material: 
 {\url{https://www.geogebra.org/m/gurqxh6x}}  .}
 \label{fig1}
\end{figure*}

Observing the phenomena of self-organization, adaptability, and cooperativity that occur during the growth of hydrogel beads as they hydrate in different containers, forming 3D, 2D, and quasi-1D structures, allows us to understand fundamental concepts of soft condensed matter. Figure \ref{fig1} shows an example of how the centers of the beads can be located. From these measurements it is possible to calculate structural quantities that characterize the degree of ordering in the system. In particular, the hexatic order parameter
\begin{equation*}
    \Psi_6 = \frac{1}{n_j}\sum_{k=1}^{n_j}\re^{6\ri\theta_{jk}}\,,
\end{equation*}
where $n_j=6$ denotes the number of neighbors of particle $j$ and $\theta_{jk}$ is the angle formed between the segment connecting particles $j$ and $k$ and a reference axis, provides a measure of how the system self-organizes as it grows.

Finally, if the container size and the particle diameter permit, the hydrogel beads suffer deformations due to compressions between them. As will be seen later, these deformations correspond to the formation of Voronoi polygons which, as the particles continue to grow, produce an adaptation of the particles to the confined space. However, to study the phenomenon of cooperativity, the use of a theoretical model is required, such as that of hard disks and their corresponding energy landscape, which can be used as long as the deformation of the hydrogel particles does not play a significant role \cite{2,3,4}.

In the present work we simulate the growth of one, two, and three hydrogel particles by using a simple elastic potential to describe the interactions between them and the container wall. Our goal is to develop a minimal and general model capable of capturing the essential mechanisms governing the collective behavior of confined hydrogel particles during growth. In particular, we aim to clarify how geometric constraints and elastic interactions jointly produce cooperative effects and metastable configurations in confined particulate systems.

In the present context we distinguish between two complementary types of interaction networks. The entropic network is defined by connectivity induced by overlaps of excluded volumes, which restrict the accessible configuration space and determine the topology of the free volume independently of mechanical deformation. By contrast, the energetic network emerges from elastic contacts between particles, and between particles and confining boundaries, storing mechanical energy upon compression. Whereas the entropic network reflects purely geometric constraints, the energetic network encodes the mechanical response of the system. The competition and mutual percolation of these two networks give rise to collective phenomena such as self-organization, adaptability, and cooperativity observed during particle growth under confinement.

From a broader perspective, the present approach is closely connected to molecular-dynamics descriptions of granular materials, in which collective phenomena emerge from purely contact-based interactions between discrete particles. In particular, pioneering molecular-dynamics studies by Gallas, Herrmann, and Soko\l{}owski \cite{4a} demonstrated how elastic repulsion, dissipation, and confinement give rise to non-trivial cooperative behavior in vibrated granular systems, including convection cells, fluidization, and pattern formation. These works established a clear link between microscopic contact mechanics and macroscopic organization in granular media, highlighting the role of elastic interactions and geometric constraints in shaping the energy landscapes and collective modes \cite{4b}. Although the present study focuses on soft hydrogel particles undergoing quasi-static growth rather than dynamic vibration, it shares the same underlying philosophy: complex collective behavior arises from simple local interactions constrained by geometry.

\noindent

The study of particle growth within a container is not new. Within the fields of liquid theories and crystallography, the first studies on sphere packing and grain growth were carried out by Stephen Hales, according to John D. Bernal, see \cite{5}. These studies described the formation of Voronoi polygons, particle contacts, and the splitting of the second oscillation of the radial distribution function. In three dimensions, this splitting characterizes the formation of disordered packings that differ significantly from ordered states. However, in two dimensions, due to the tendency toward crystallization, this splitting describes the formation of crystalline nuclei through the formation of cages produced by overlapping excluded volumes \cite{6}. Similar confinement-induced ordering phenomena have also been studied in quasi-one-dimensional systems \cite{7,8}. In such systems the overlaps of excluded volumes produce extended connectivity throughout the system, which in the present work is interpreted as the formation of entropic networks.

On the other hand, and related to the phenomena of cooperativity, there are memory effects that include the history of how the materials were made. Examples include hysteresis observed in magnetic systems and the appearance of interfaces separating condensed phases in the presence of phase coexistence. In such systems cooperative effects can lead to metastable states that retain information about the path through which they were obtained. Recently, Hagh and colleagues from Nagel's group reported a spring model whose bistability is responsible for such memory effects in the so-called ``hysterions'' \cite{10}. This might resemble the case of hydrogel beads because it is also an elastic model. However, although the growth of hydrogel beads produces elastic deformations when particles contact each other or the confining wall, these contacts cannot elongate, unlike springs. Therefore, the model describing these interactions must respond elastically to compression rather than to separation.

Motivated by these considerations, in this work we introduce a simple elastic model for confined hydrogel particles undergoing hydration-induced growth. The model is intentionally formulated in a dimensionless and general form, where the elastic constant is set to $k=1$ without loss of generality. This choice permits the model to capture universal geometric and mechanical aspects of confined particle growth independently of the specific material properties of a given hydrogel system. Using this minimal framework, we investigate how elastic contacts and excluded-volume constraints combine to produce cooperative structural evolution in systems containing one, two, three, and several particles confined within a circular boundary.

\section{Modelling and methodology}\label{sec2}

To construct the proposed elastic model, we assume that a hydrogel particle undergoes linear radial growth upon hydration. The system considered consists of $N$ circular particles confined within a circular container. The particles increase their radius progressively during hydration while remaining free to move within the container in order to minimize the elastic energy generated by contacts with other particles or with the container wall.

When a particle of radius $r$ contacts the wall of the container, an interaction potential $u(\Delta r)$ is established
\begin{equation}
  u(\Delta r)=\frac{1}{2}k(\Delta r)^2\,,  
\end{equation}
where $k$ is the elastic constant and $\Delta r$ represents the radial deformation of the particle. In the present model we set $k=1$. This choice corresponds to a convenient nondimensionalization of the elastic energy and permits the model to be formulated in a general form independent of the specific elastic properties of a given hydrogel material. The parameter $\Delta r$ represents the overlap distance produced by compression. In the case of particle-wall interaction it is defined as
\begin{equation}
    \Delta r = r + d - R\,,
\end{equation}
where $R$ is the radius of the circular container and $d$ is the distance between the particle center and the center of the container. Similarly, for two particles $i$ and $j$ separated by a center-center distance $d_{ij}$, the deformation is given by
\begin{equation}
\Delta r_{ij} = r_i + r_j - d_{ij}\,.    
\end{equation}

The elastic interaction is activated only when $\Delta r > 0$, that is, when particles or particle and wall are in contact and compression occurs. This potential drives the particles toward configurations corresponding to local minima of the total elastic energy.

\begin{figure*}[!ht]
\includegraphics[width=0.30\textwidth]{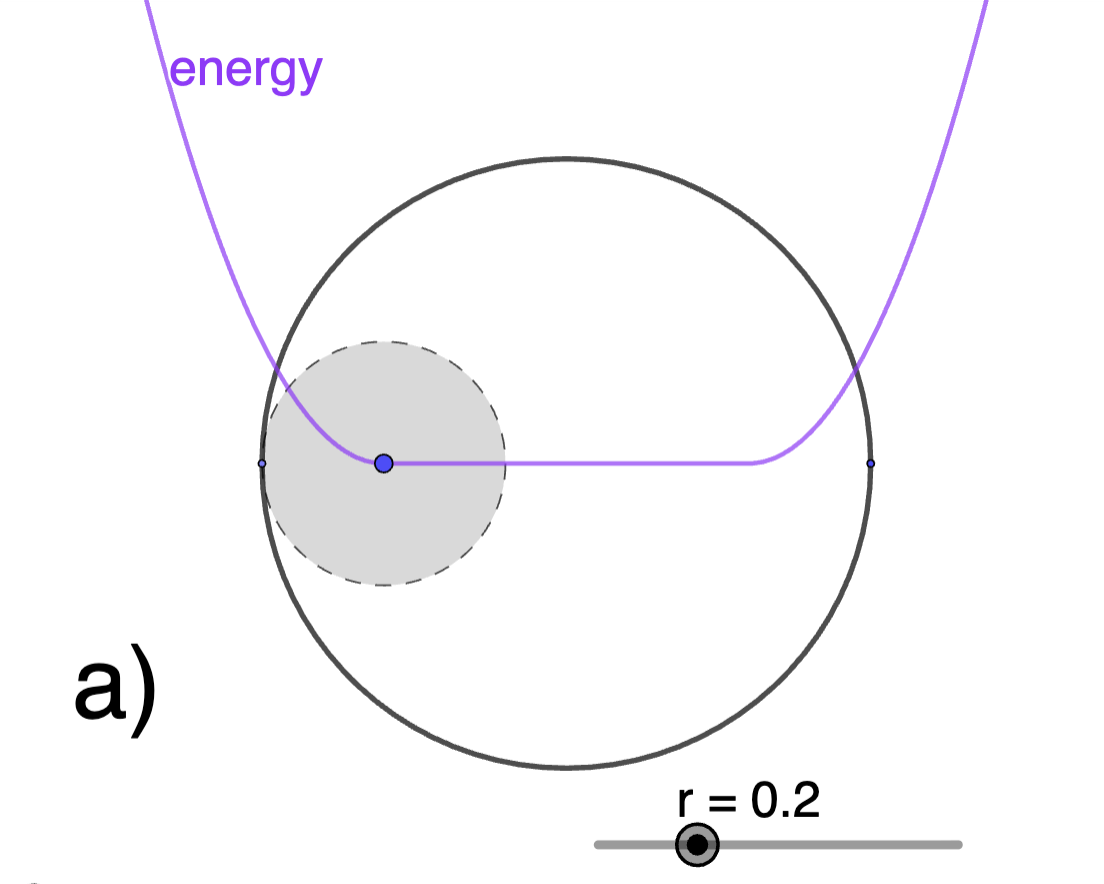}
\includegraphics[width=0.30\textwidth]{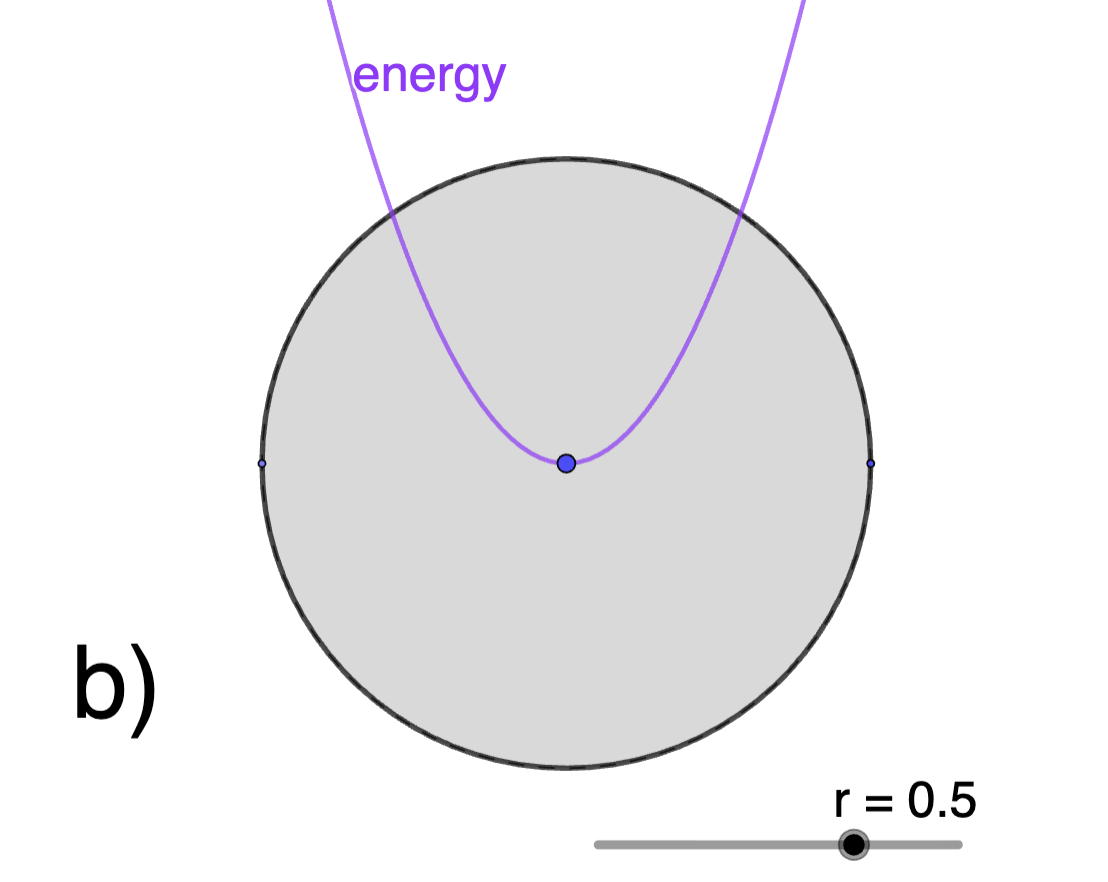}
\includegraphics[width=0.30\textwidth]{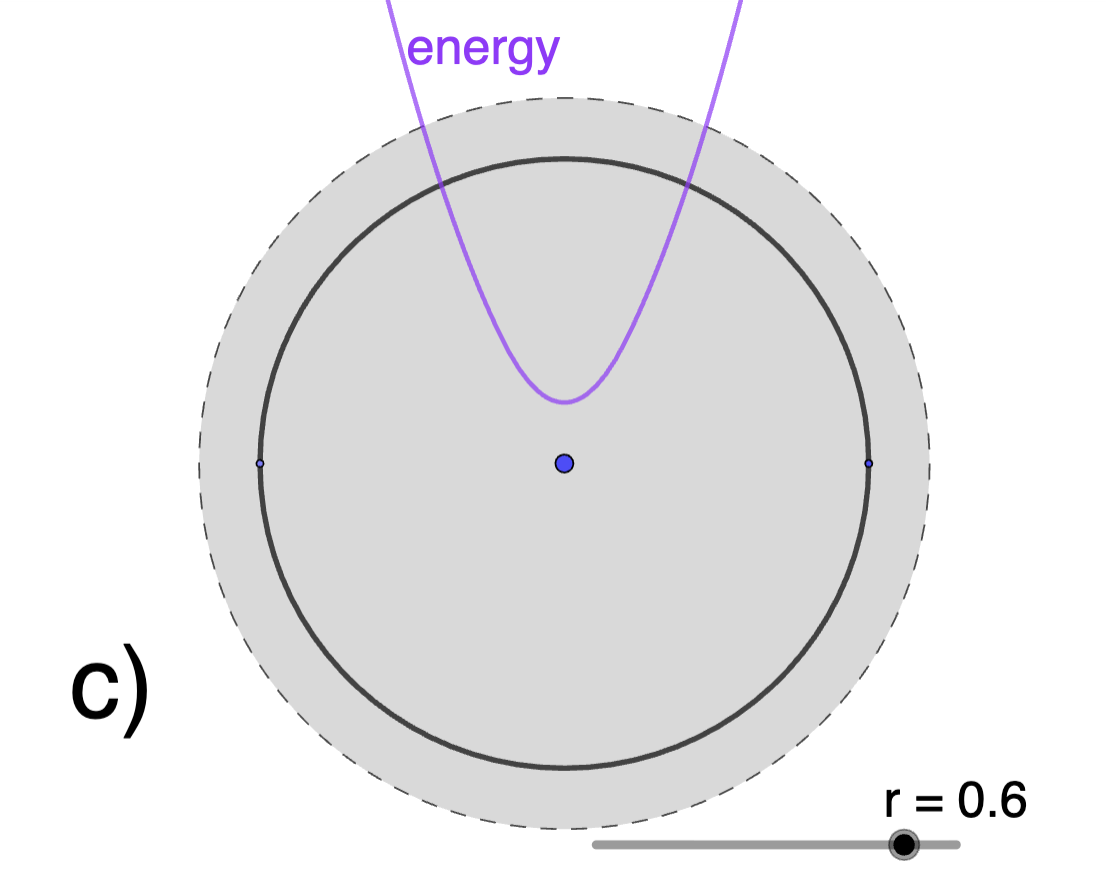}
 \caption{(Colour online) Model of the displacement dynamics of a hydrogel particle in a circular container, whose center is located at the minimum of the potential energy landscape, considering ``Hookean'' contacts upon reaching the wall (i.e., the simplest case of percolation) during growth. (a) The particle first touches the circular wall and begins to move, producing a displacement of its center toward the center of the container. (b) The particle becomes comparable in size to the container and touches the wall over an extended region as equilibrium is approached. (c) Further growth leads to the accumulation of elastic potential energy. An animation for a single particle interacting with the circular container is provided as supplementary material: {\url{https://www.geogebra.org/m/hv9avwhh}}.}
 \label{fig2}
\end{figure*}

Figure \ref{fig2} illustrates the simplest case of this process for a single particle interacting with the circular boundary. As the particle grows, it eventually touches the wall and begins to move, producing a displacement of its center toward the center of the container. As the growth continues, the particle becomes comparable in size to the container and the contact region increases, resulting in the accumulation of elastic potential energy.

\noindent

\begin{figure*}[!ht]
\includegraphics[width=0.245\textwidth]{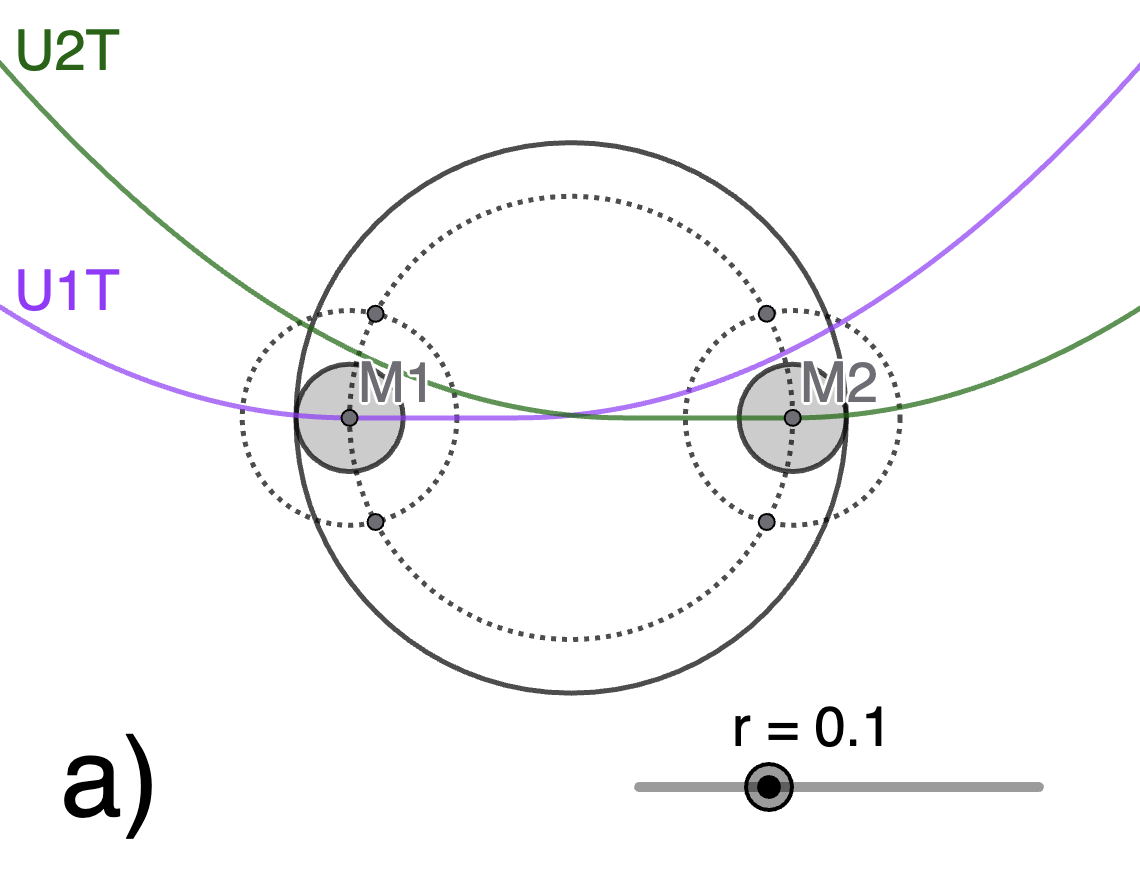}\includegraphics[width=0.245\textwidth]{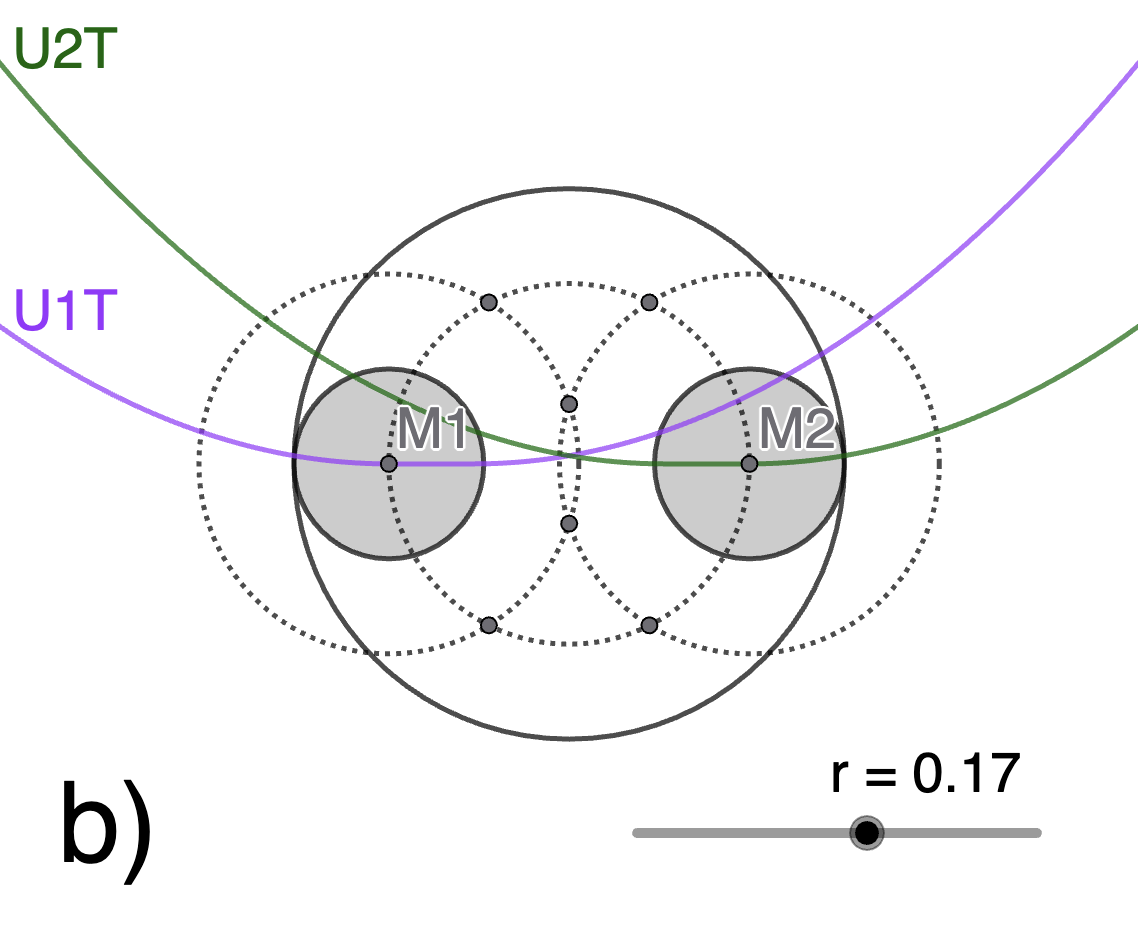}\includegraphics[width=0.245\textwidth]{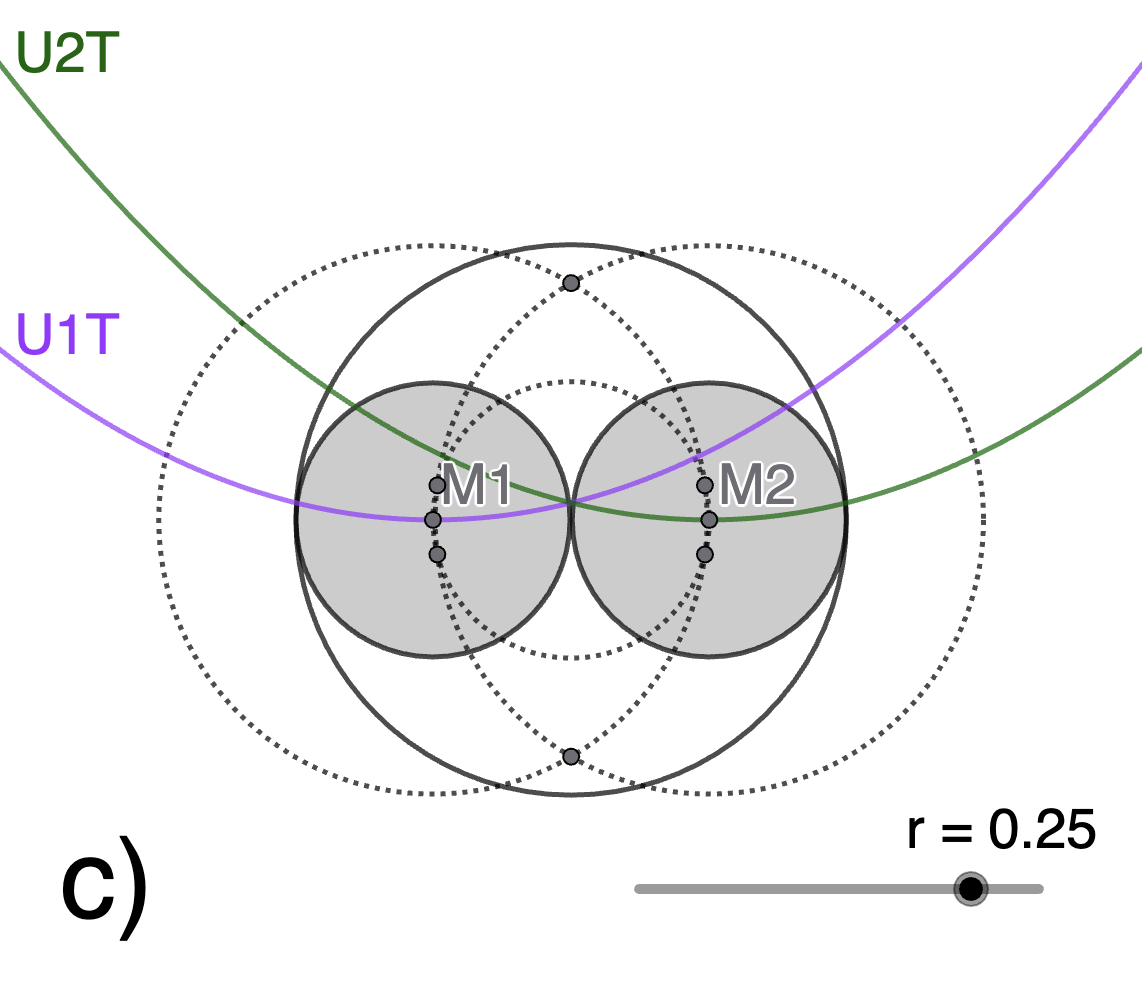}
\includegraphics[width=0.245\textwidth]{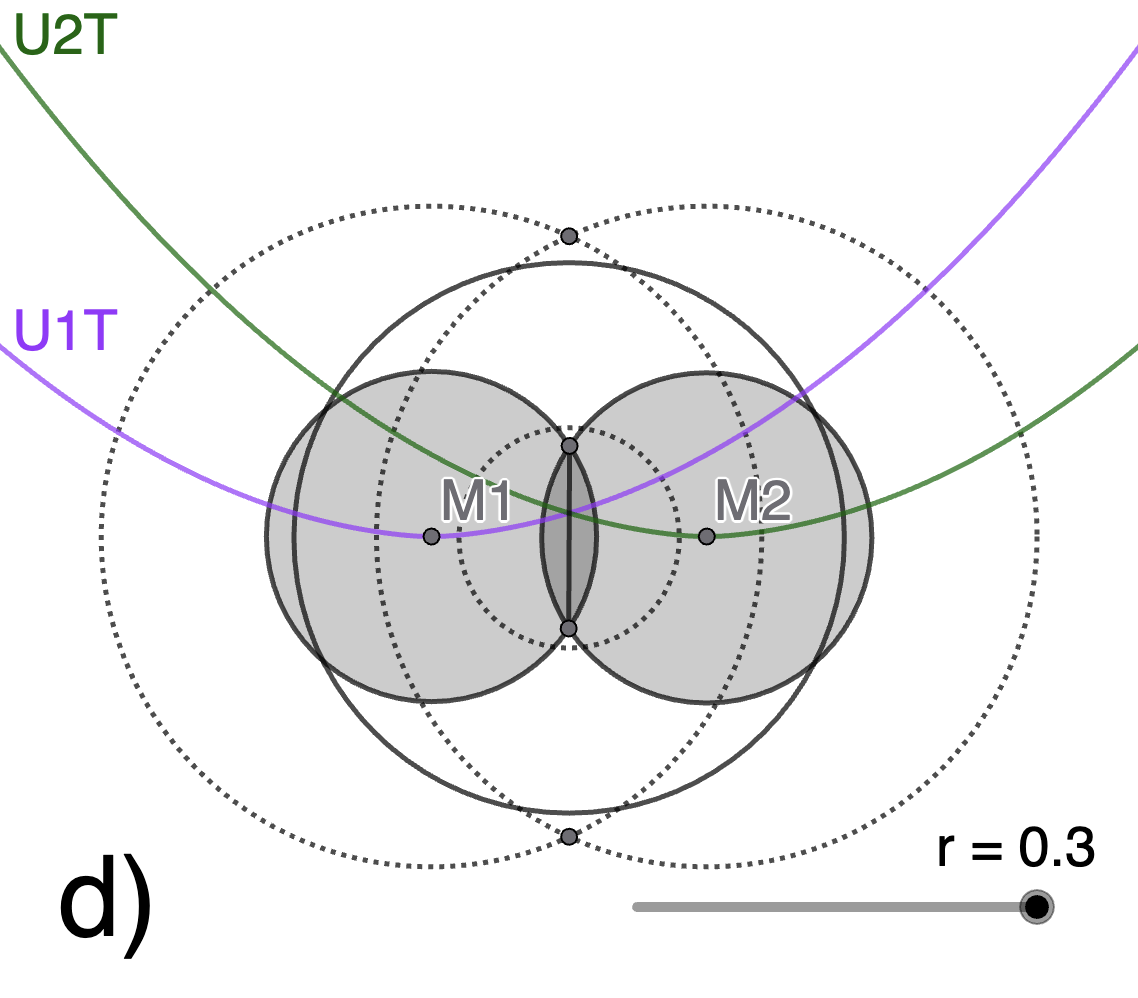}
 \caption{(Colour online) Modelling the displacement dynamics of two hydrogel particles within a circular boundary during hydration-induced growth. The particle centers move toward the minimum of the potential energy landscape observed by each particle. We show the excluded volume for the growing particles and container-particles, a) shows overlap of the excluded volume with the container; b) shows the case when the overlap of the excluded volume between particles occurs; c) when the particles starts touching each other; and d) the two-particle system starts to accumulate elastic energy and adapts to particle-particle and particle-wall interactions. As mentioned before, notice the importance of contact percolation between walls to achieve equilibrium. As before, we provide a link to show an animation of both particles growing dynamics as supplementary material: {\url {https://www.geogebra.org/m/kq4caksx}}  .}
 \label{fig3}
\end{figure*}

In the two-particle system (figure \ref{fig3}), the particles are initially placed at opposite sides of the container. As they expand they eventually interact elastically with each other and with the confining boundary. Each contact contributes an energy term of the form
\begin{equation}
   u_i(\Delta r_i)=\frac{1}{2}k(\Delta r_i)^2\,. 
\end{equation}

\noindent

Since the interaction between two particles can be interpreted as two elastic elements acting in series, the equivalent elastic constant $k_{\text{eq}}$ satisfies
\begin{equation}
    \frac{1}{k_{\text{eq}}}=\frac{1}{k}+\frac{1}{k}\,,
\end{equation}
which for $k=1$ yields $k_{\text{eq}}=0.5$. Although the present work uses $k=1$, this parameter can be rescaled to match specific experimental systems if necessary.

Figure \ref{fig3} (a-d) illustrates the excluded volume (based on particle diameter) in four distinct stages. Whereas panel (b) shows the overlap of excluded volumes, panel (d) depicts the actual volumes and the remaining free volume, highlighting the capacity of the particles to adapt to spatial constraints through deformation. The overlapping volume at the moment of percolation signifies the storage of elastic potential energy within the system. 

\begin{figure}[h]
\includegraphics[width=0.245\textwidth]{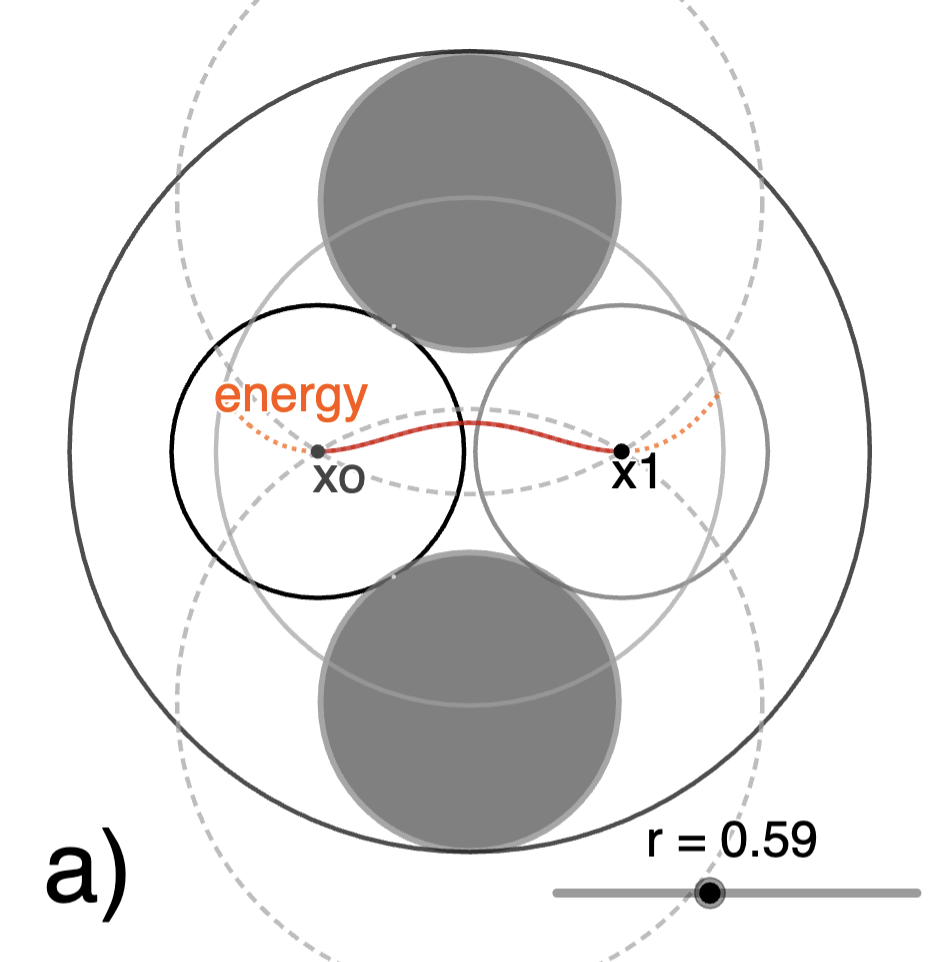}\includegraphics[width=0.25\textwidth]{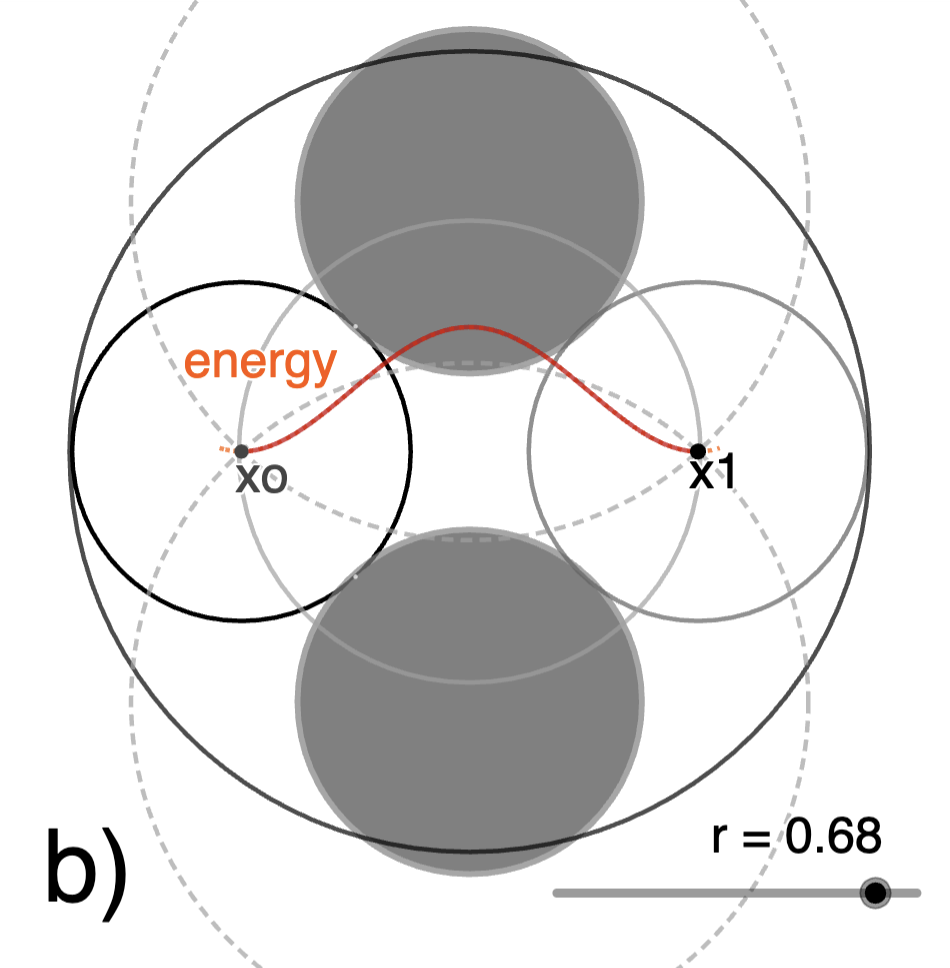}
\includegraphics[width=0.245\textwidth]{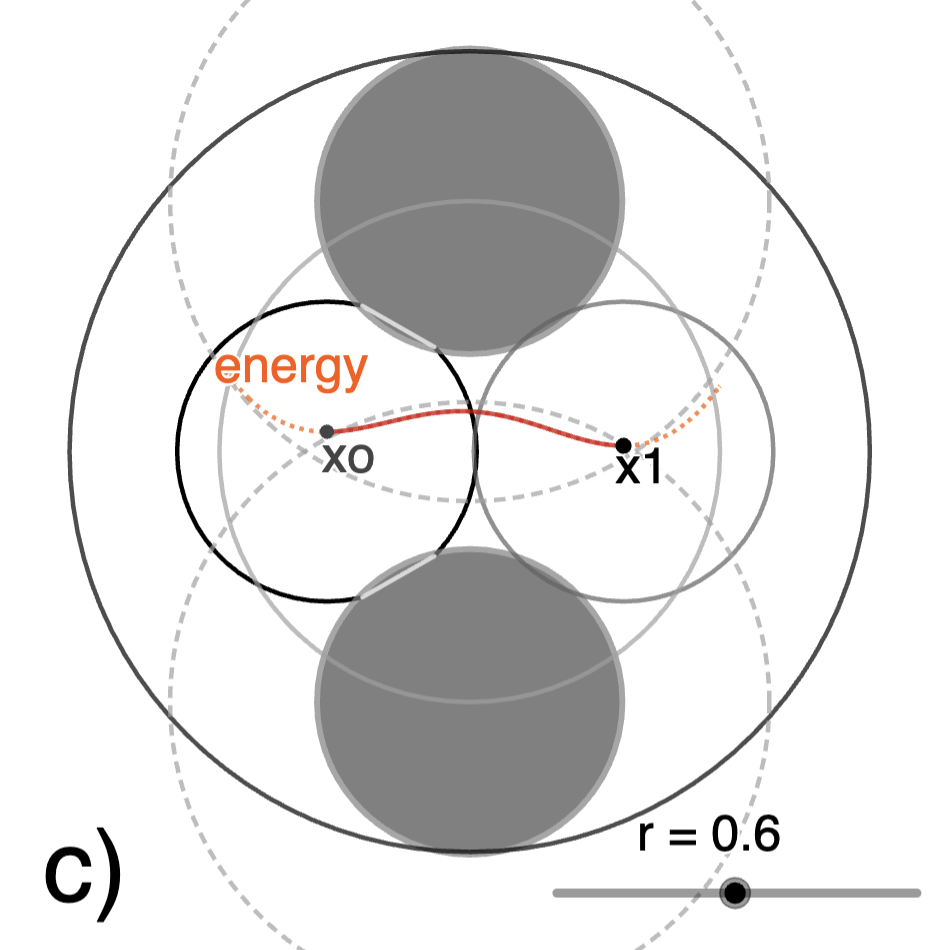}\includegraphics[width=0.245\textwidth]{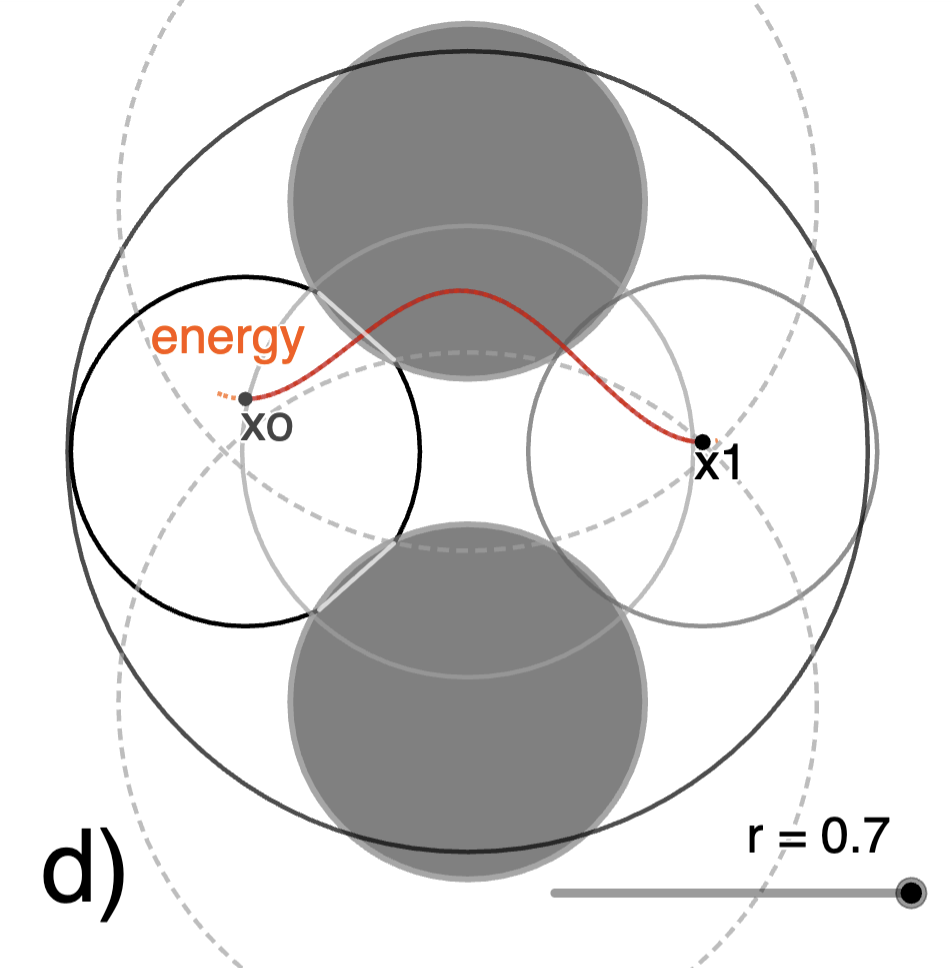}
 \caption{Modelling the displacement of a particle placed between other two particles during its growth (sequence a-b), the center of the free particle tracks the minimum of the potential energy landscape produced. In c-d), the application of a lateral external force induces the emergence of a local (metastable) minimum. Unlike spring-based models (10), only the repulsive effect (indicated in red) is important. The case of three particles interacting with each other and an external force, which can be considered as another particle, a set of particles, and/or a wall, as appropriate:  
 {\url {https://www.geogebra.org/m/dye5x3ph}} .}
 \label{fig4}
\end{figure}

Figure \ref{fig4} depicts a three-particle system. Two particles (indicated in dark gray) are fixed vertically, while the energy of a third, mobile particle is plotted (red line) as a function of its center's position. This energy profile reveals a potential barrier that the mobile particle must overcome to transition across the excluded volume overlap. Although panels a-b show unconstrained growth, panels c-d demonstrate the effect of an external force applied from the left. This force shifts the potential energy landscape, raising the minimum on the left side of the barrier and creating a metastable state (a local minimum). The white lines in these panels denote the deformations resulting from the volume overlap under this external load. Across all configurations in figure \ref{fig4}, a ``buckling'' phenomenon is observed. This was initially suggested in the two-particle case, where two regions of free volume emerge perpendicular to the axis that join the particle centers (figure \ref{fig3}b). In the hard-disk limit, where particle deformation is neglected, the system converges toward the zero-temperature energy landscape proposed by Hunter and Weeks \cite{4}.

\textbf{Numerical implementation.} 
The evolution of the system during growth is modelled through a quasi-static procedure in which the particle radii are increased gradually while the particle positions are continuously adjusted to minimize the total elastic energy of the system. At each growth step, the radius of every particle is increased by a small increment
\begin{equation}
    r(t+1)=r(t)+\delta r\,,
\end{equation}
where $\delta r$ represents the hydration-induced growth step.

After each increment of the particle radii, the positions of the particle centers are permitted to relax by minimizing the total elastic energy
\begin{equation}
    U=\sum_{\rm contacts}\frac{1}{2}(\Delta r)^2 \,.
\label{Uelast}
\end{equation}

The minimization process is implemented numerically using an iterative relaxation procedure that moves the particle centers toward the local minimum of the potential energy landscape. The resulting configuration represents mechanical equilibrium for that growth step.

This procedure reflects the quasi-static nature of hydrogel hydration: the particles expand slowly compared to the time required for mechanical relaxation of their positions. As a consequence, the system can be treated as evolving through a sequence of equilibrium configurations.

\textbf{Initial configurations.} 
For the systems studied in this work, the initial particle positions are selected such that no overlaps exist between particles or between particles and the container walls. In the case of one or two particles, symmetric initial configurations are used. For systems with a larger number of particles, the particles are initially placed inside the container in configurations that regard the excluded-volume condition and permit the subsequent growth dynamics to determine the final structure.

\textbf{Physical interpretation.} 
During the early stages of growth, the particles do not touch each other and their motion is governed primarily by geometric constraints associated with excluded volumes. These constraints define what we call the entropic network, which describes the connectivity of accessible free volume regions. As growth proceeds and particles come into contact with one another or with the confining boundary, elastic compression occurs and elastic energy begins to accumulate. These contacts form what we refer to as the energetic network, which represents the subset of connections that actively transmit the mechanical stress.

The interaction between these two networks --- one defined by geometric constraints and the other by elastic contacts --- determines the cooperative structural evolution observed in the system. As will be shown in the following sections, even in the simplest case of three particles, this competition produces energy barriers and metastable states that are characteristic signatures of cooperative behavior.

\noindent

\begin{figure*}[!ht]
\includegraphics[width=0.25\textwidth]{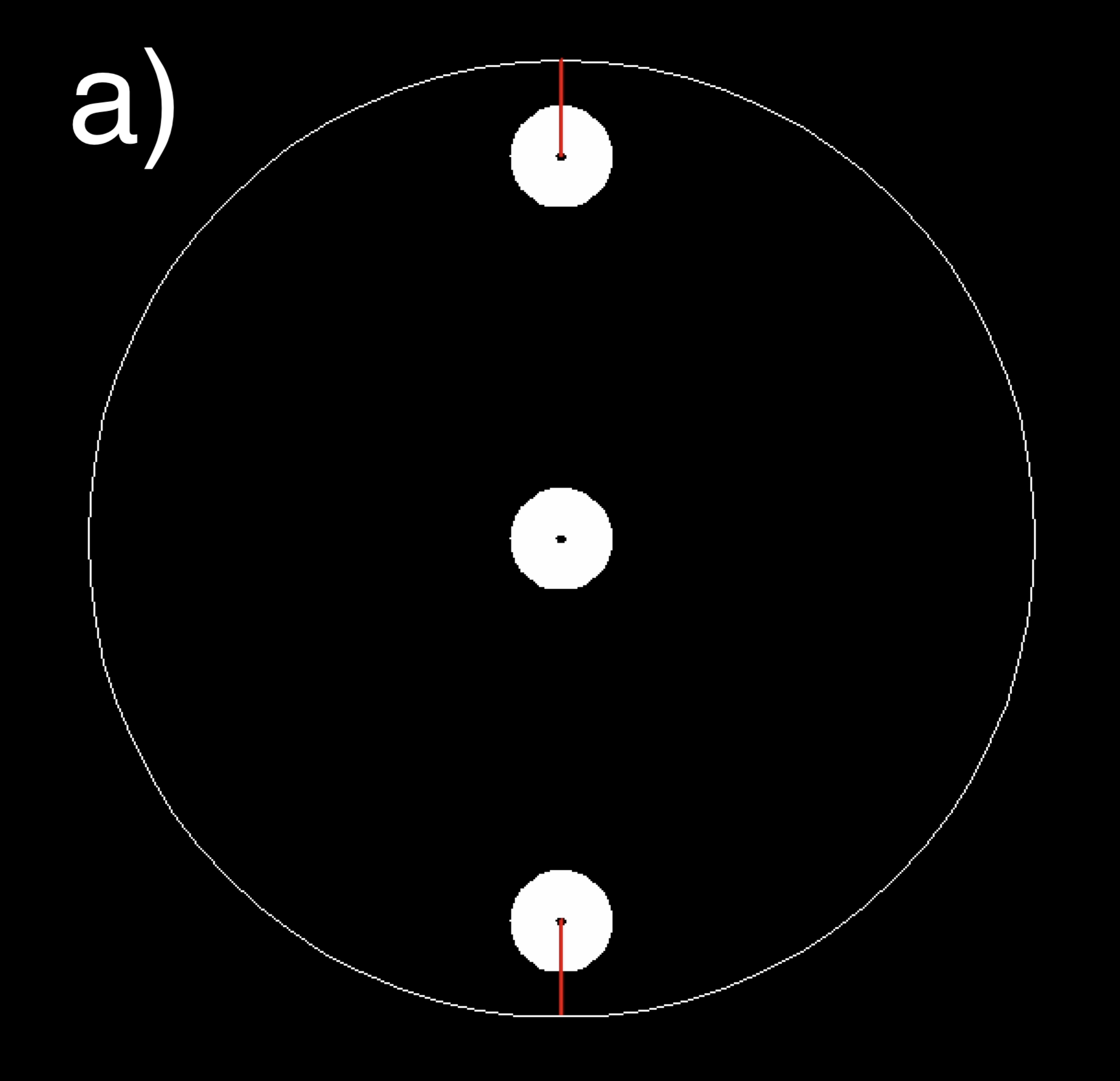}\includegraphics[width=0.25\textwidth]{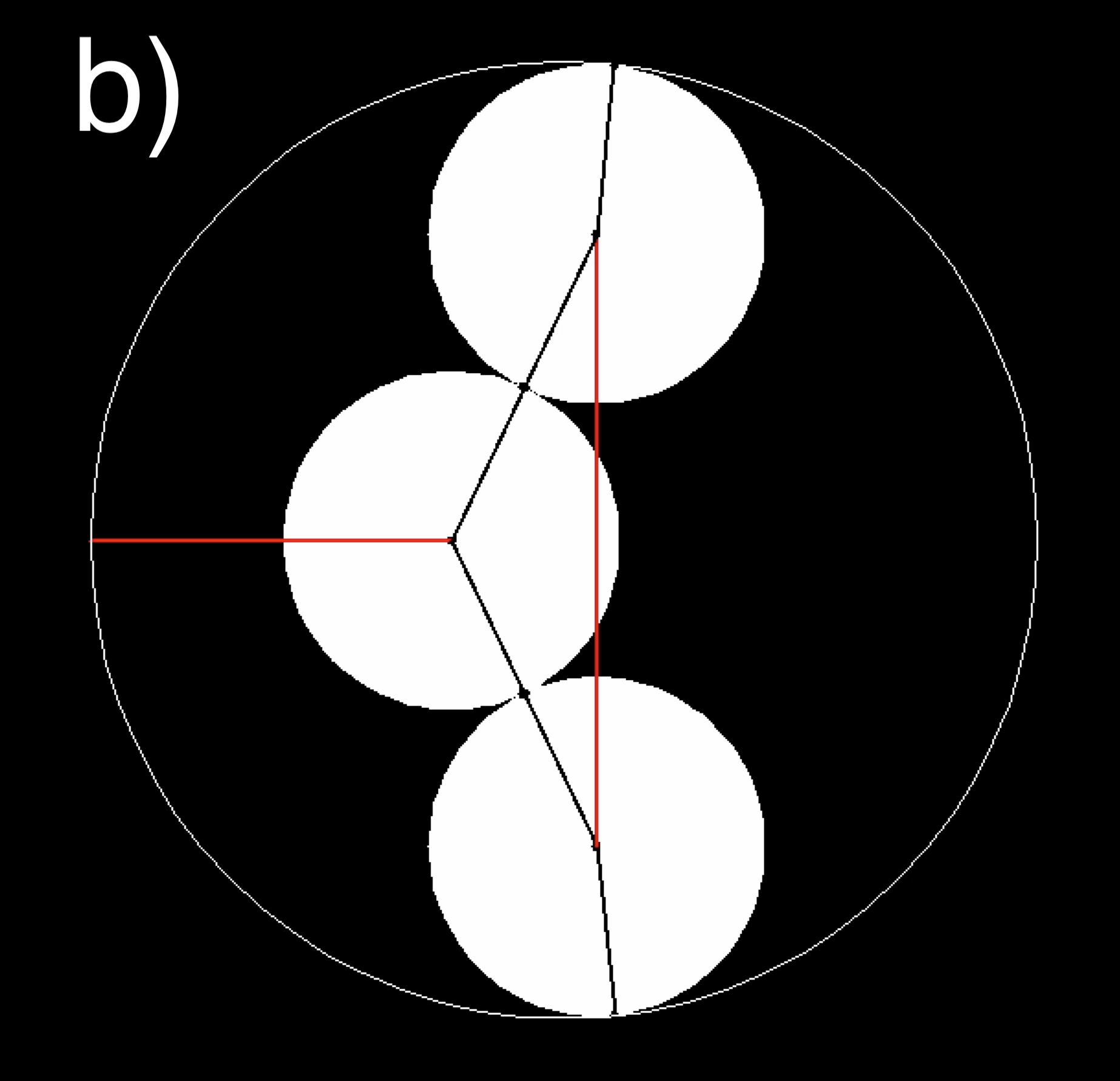}
\includegraphics[width=0.25\textwidth]{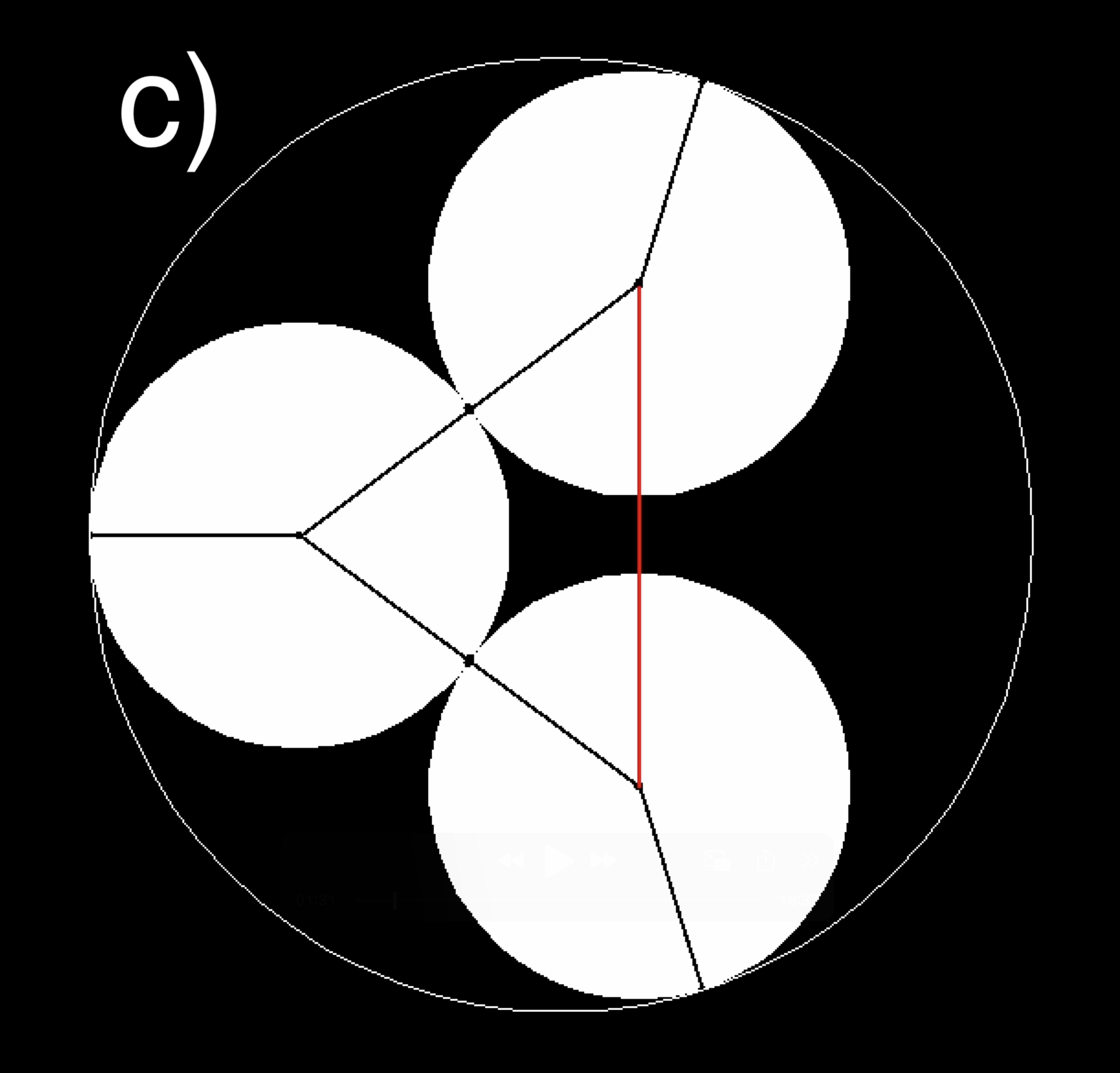}\includegraphics[width=0.25\textwidth]{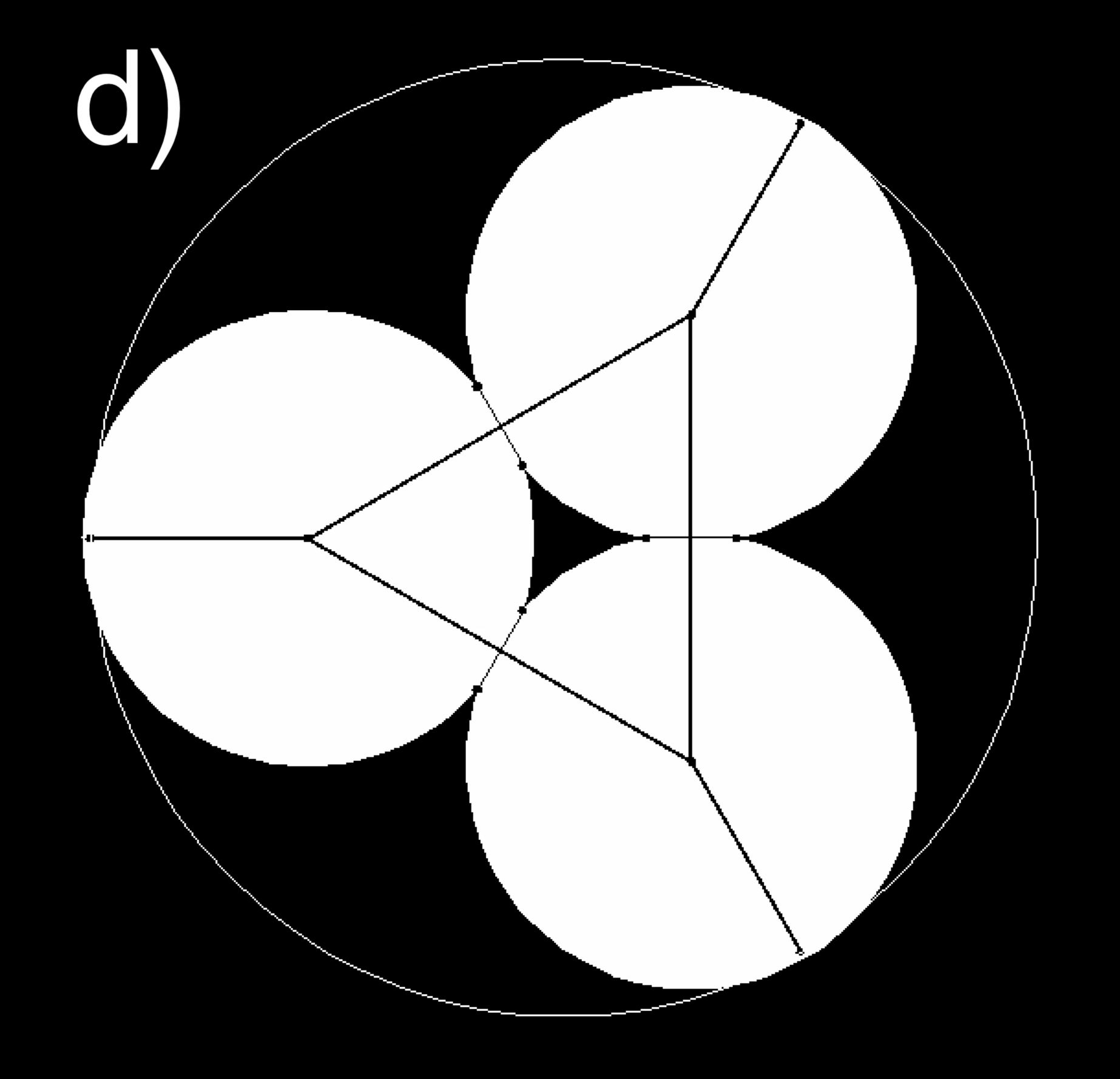}
\includegraphics[width=0.25\textwidth]{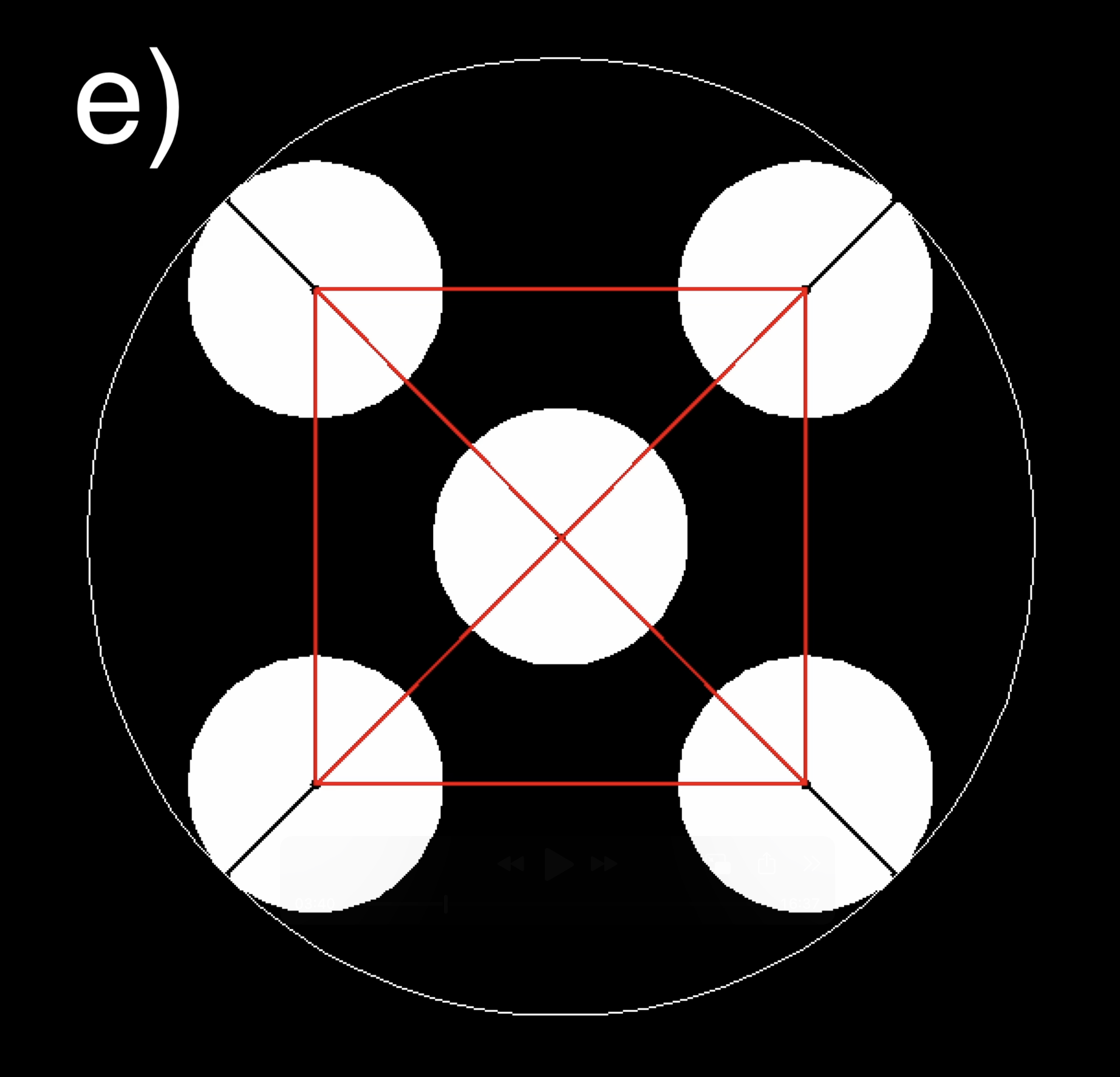}\includegraphics[width=0.25\textwidth]{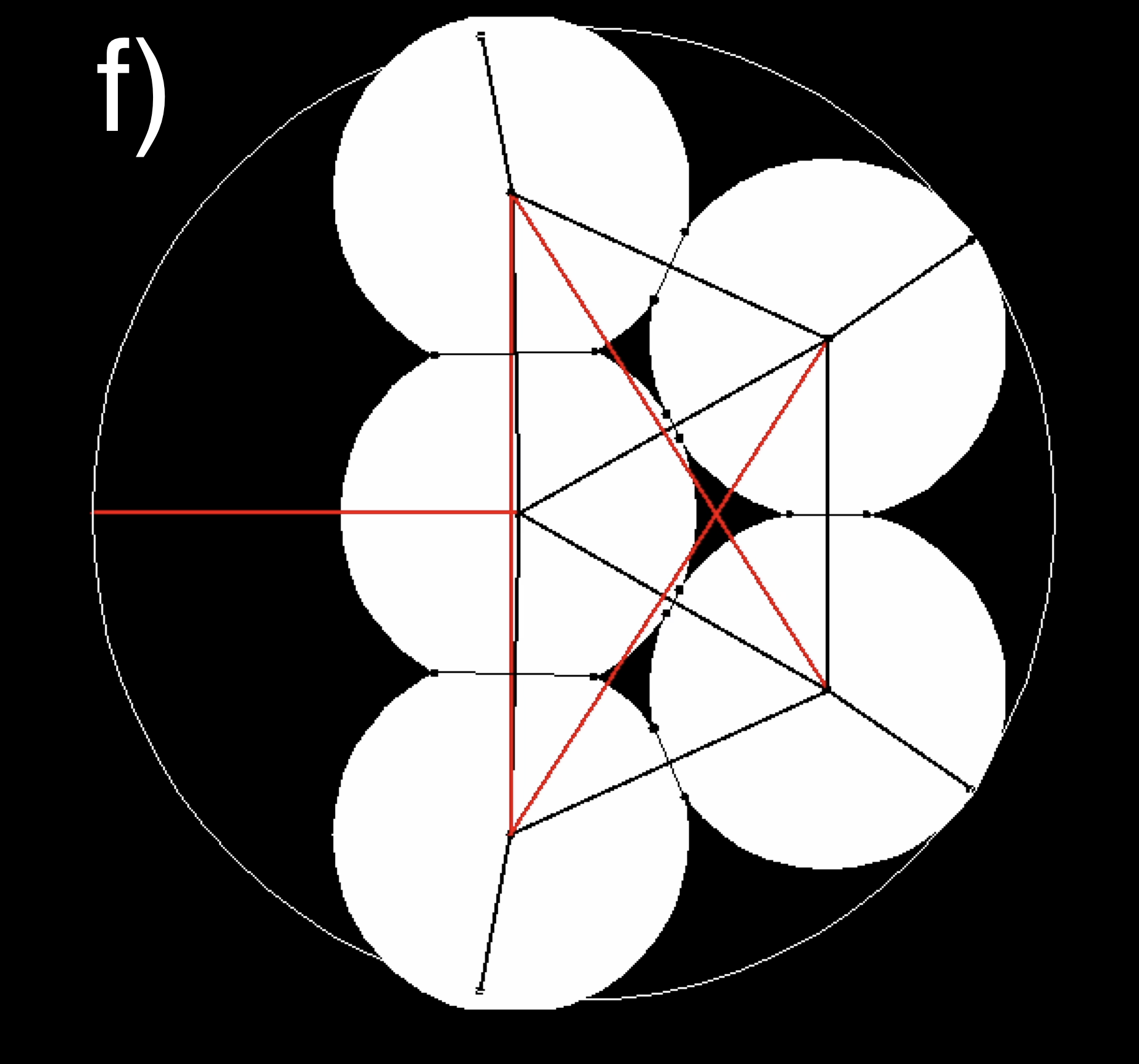}
\includegraphics[width=0.25\textwidth]{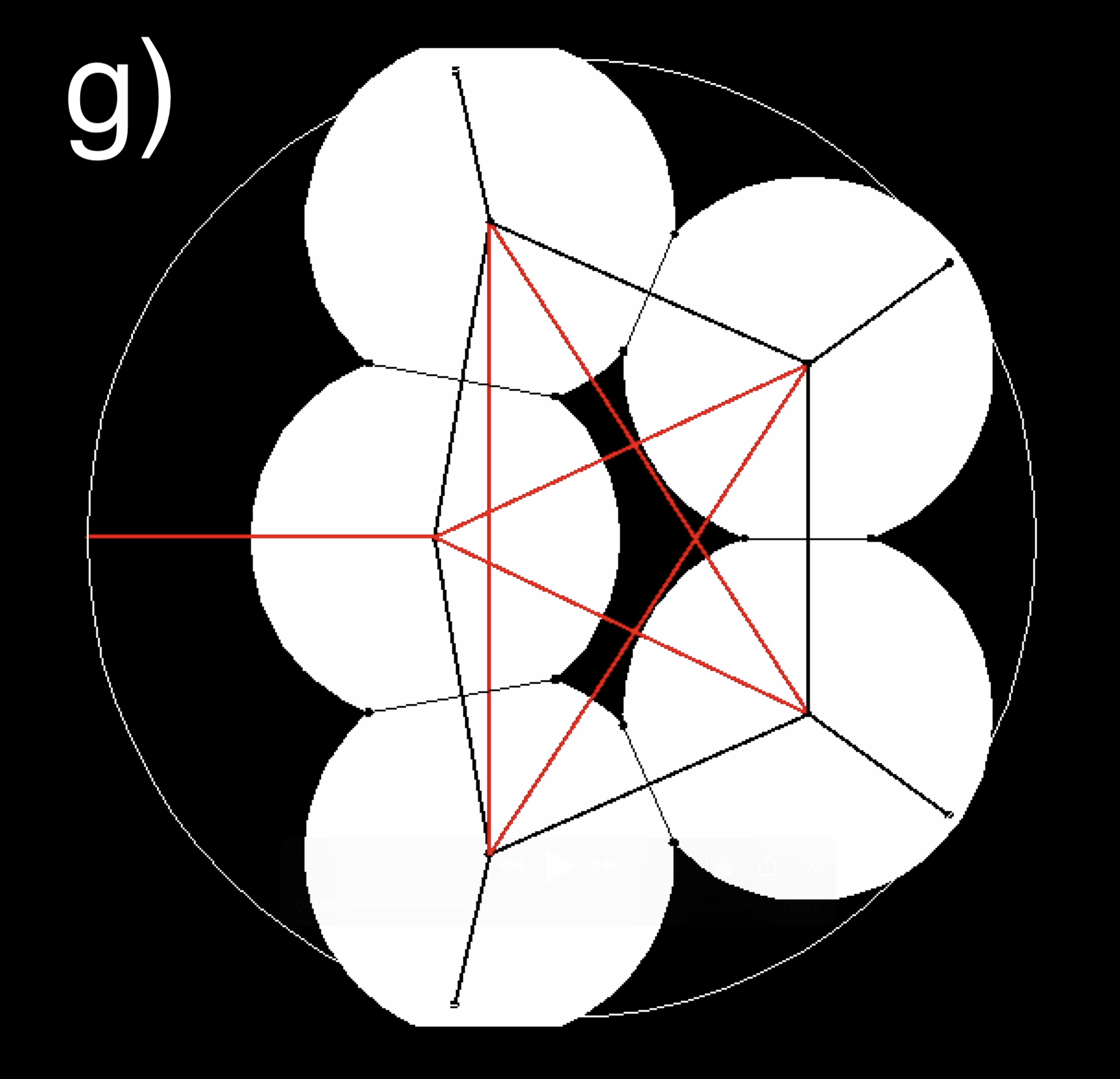}\includegraphics[width=0.25\textwidth]{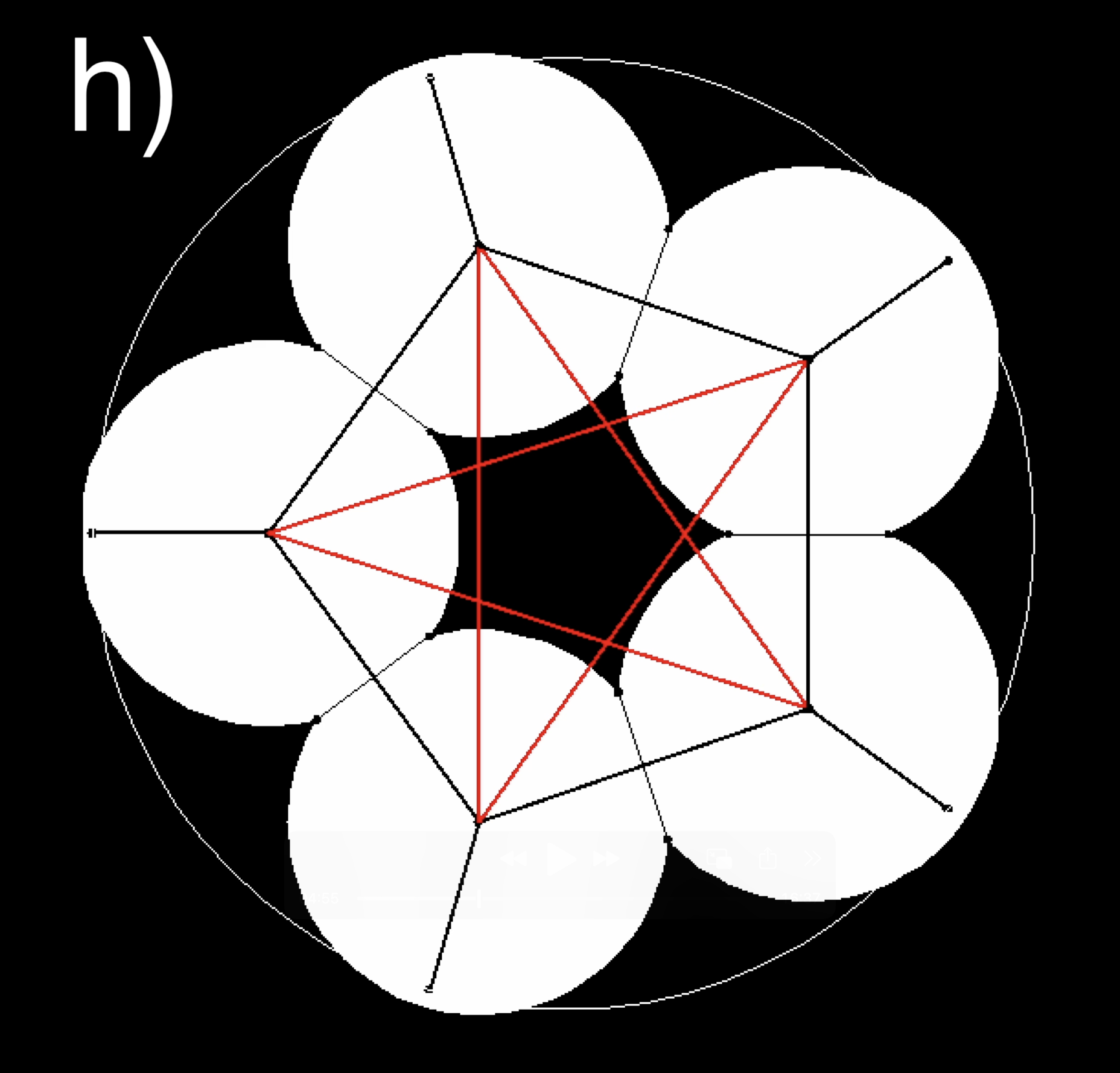}
\includegraphics[width=0.25\textwidth]{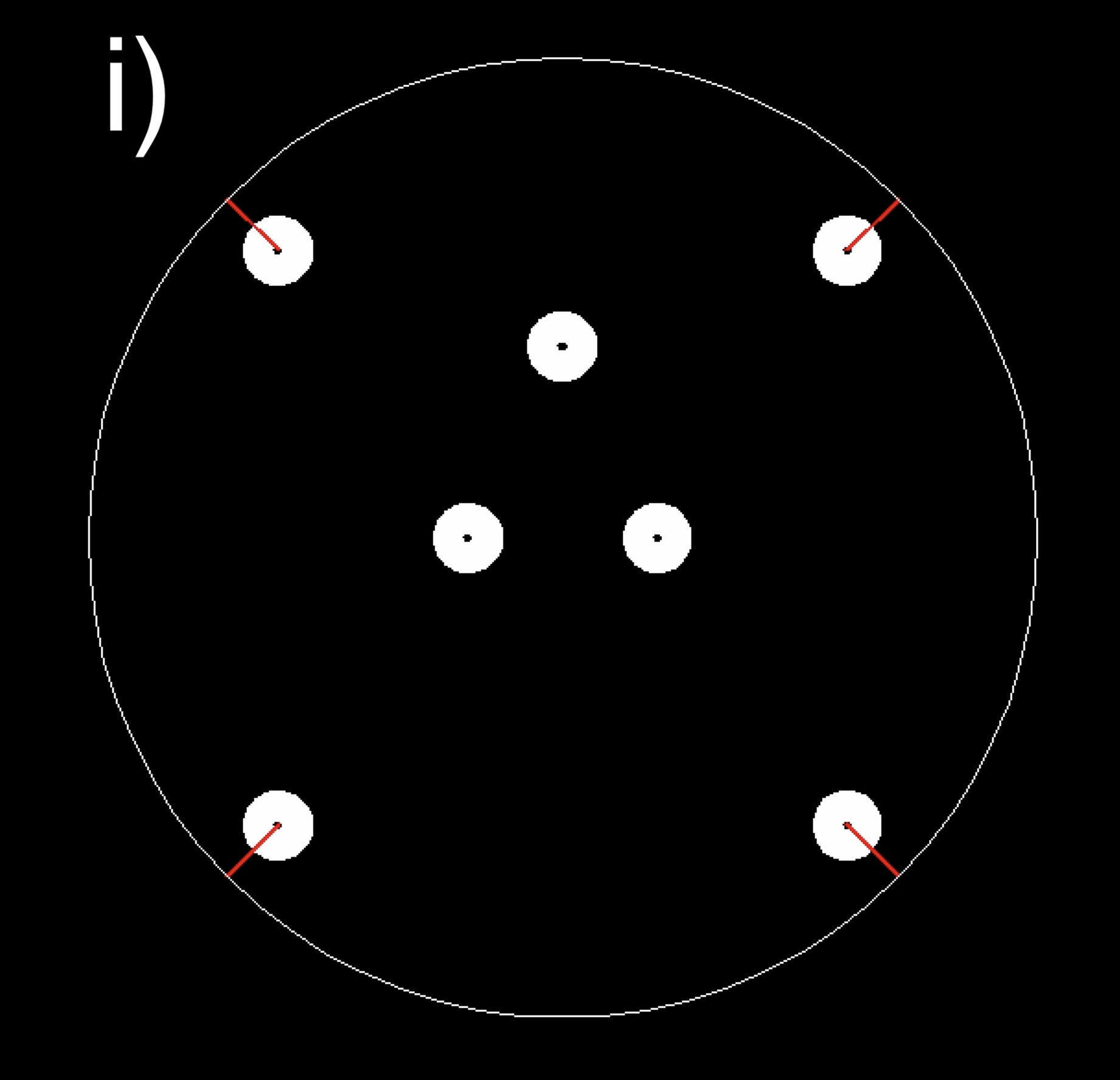}\includegraphics[width=0.25\textwidth]{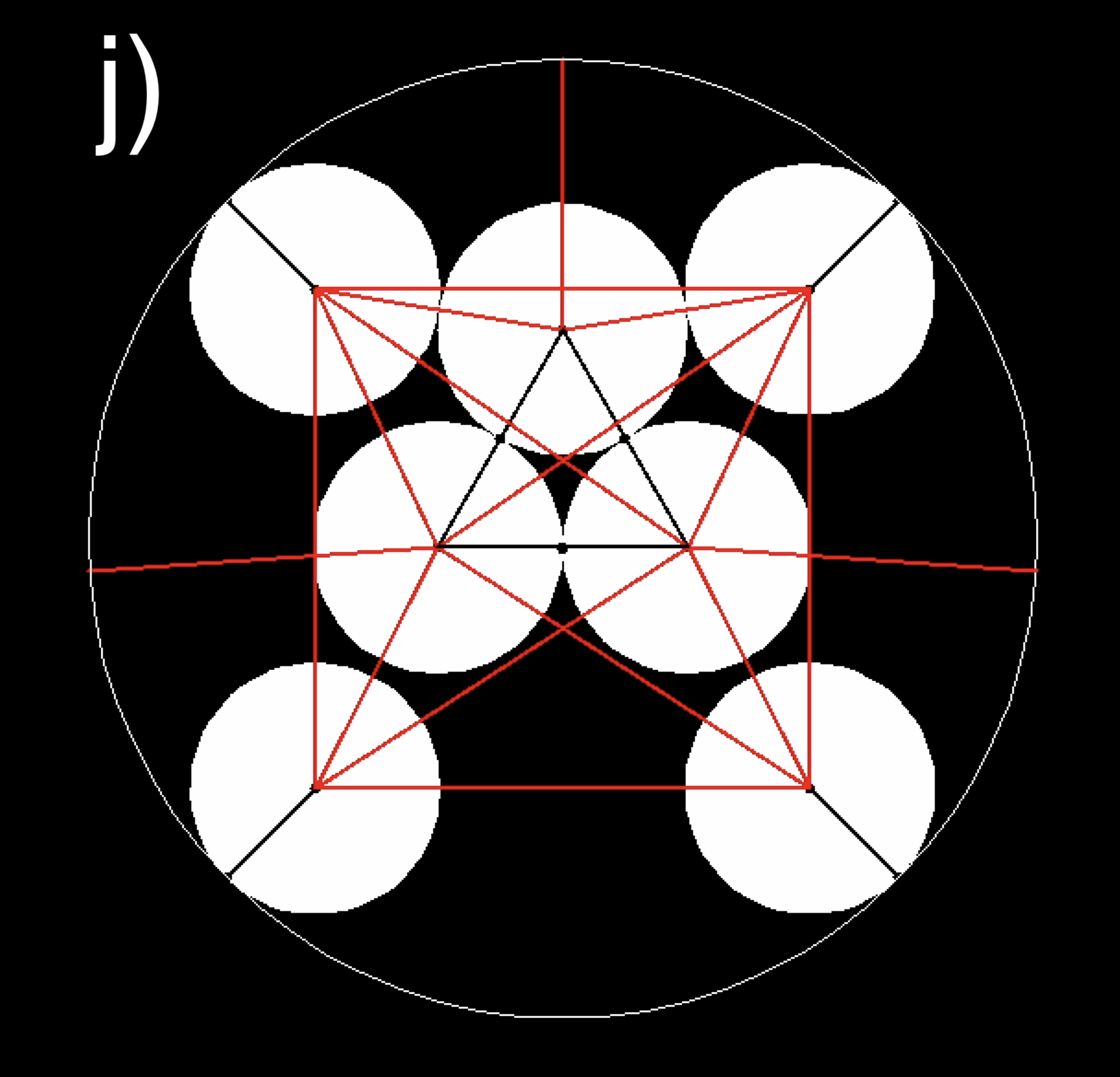}
\includegraphics[width=0.25\textwidth]{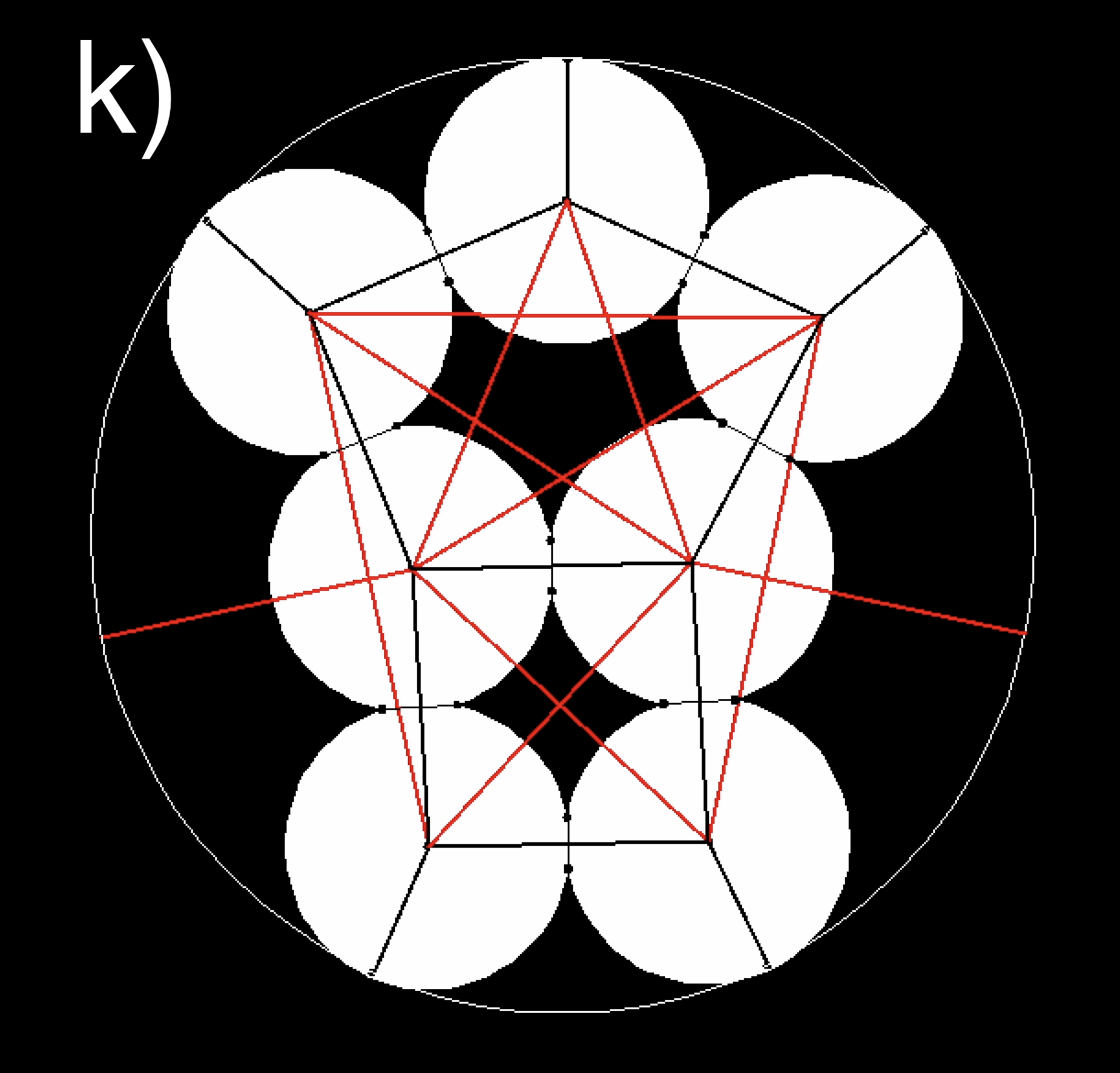}\includegraphics[width=0.25\textwidth]{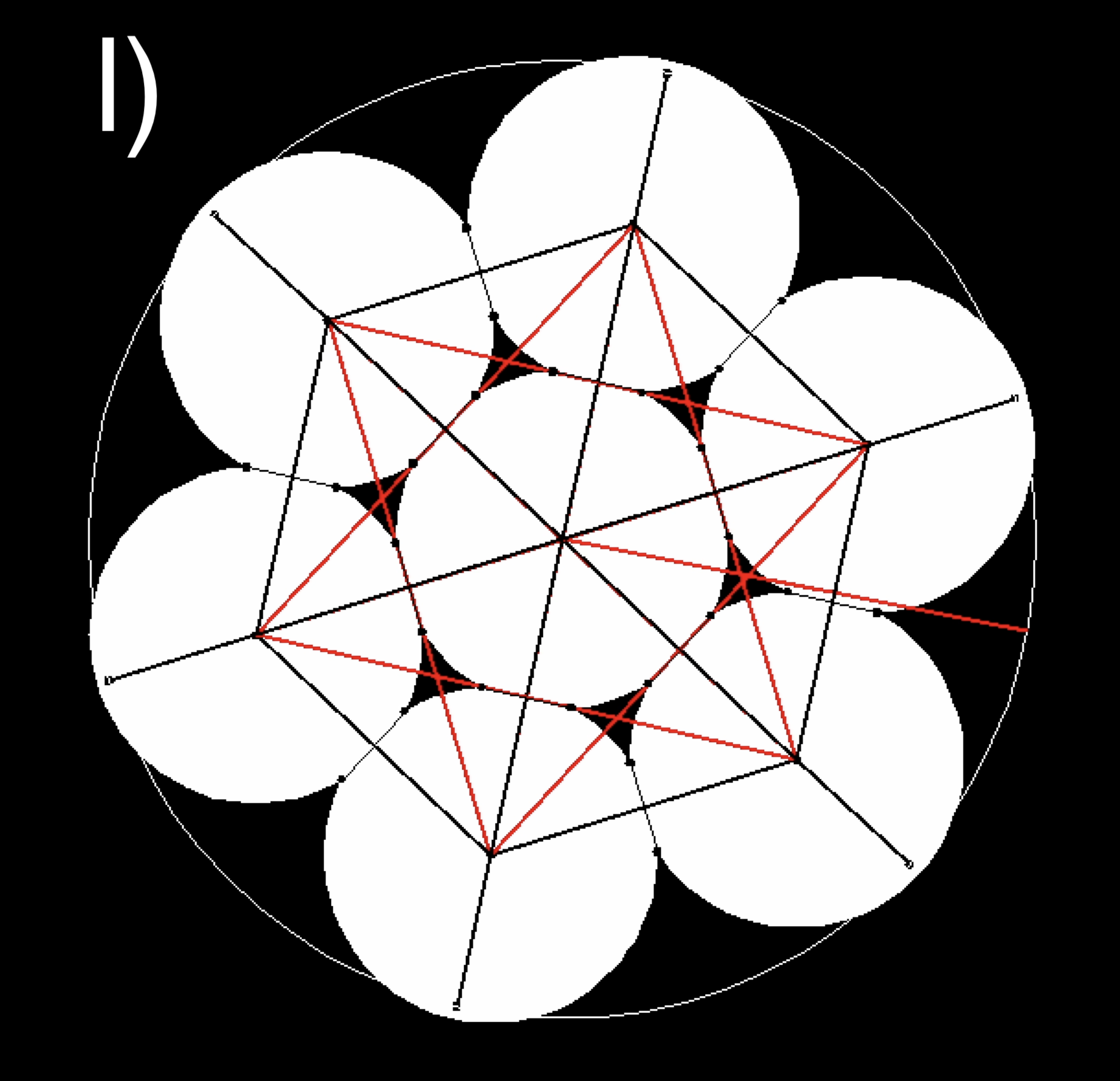}
\caption{(Colour online) Sequence of configurations obtained with the elastic model described in the text during the growth of a-d) three particles; e-h) five particles; i-l) seven particles. The following supplementary videos provide further details on these processes: 1) Simulation Overview: Demonstrates various cases used to construct the elastic model of hydrogel particles through energy minimization, featuring simulations for one, two, three, five, and seven particles. (\url{https://youtu.be/BChVR5gjIGg}); 2) Six Particles Comparison: It compares the cases involving six particles studied for the elastic hydrogel model using the energy minimization process (\url{https://youtu.be/cwFcXFLW28M}) .}
 \label{fig5}
\end{figure*}
\noindent

\section{Results and discussion}\label{sec3}

Figure \ref{fig5} illustrates the results of the model implemented via FORTRAN, using the pgplot5 library for visualization. The first, second, and third rows display cases involving three (a-d), five (e-h), and seven (i-l) particles, respectively, each evolving from the initial configurations shown in panels a, e, and i. During the growth phase, an entropic network is formed from overlapping excluded volumes, represented by red lines. These overlaps effectively restrict the particle motion by defining the connectivity of the free volume.

As the particles expand and make contact with each other or with the walls of the container, the overlaps are progressively transformed into physical contacts (indicated by black lines). These contacts generate elastic compression and therefore contribute to the total elastic energy of the system according to the potential introduced in section~\ref{sec2}. In this way, an energetic network emerges as a subset of the pre-existing entropic network.

As particle growth proceeds, new contacts appear and the elastic energy of the system evolves accordingly. The total elastic energy of the system defined in equation~\ref{Uelast} can therefore be written as
\begin{equation}
    U = \sum_{\rm contacts} \frac{1}{2} (\Delta r)^2 \,,
\nonumber
\end{equation}
where the deformation $\Delta r$ represents the compression produced by each contact. As the particles grow, the system evolves through a sequence of quasi-static configurations that correspond to local minima of this energy landscape. Since the particle radii increase monotonously during hydration, the elastic energy of the system generally increases during growth, except when structural rearrangements occur that permit the system to reach a new local minimum of the energy landscape.

This energetic network competes for the limited volume restricted by excluded-volume overlaps, resulting in a cooperative effect that may maintain the system in metastable states. Such a situation is illustrated in figure \ref{fig4} (c-d), where the potential energy profile of a mobile particle placed between two fixed particles exhibits the appearance of a local minimum separated by an energy barrier. In this configuration, the particle can remain temporarily trapped in a metastable state until a sufficiently large perturbation permits the system to cross the barrier.

The stability of these configurations arises from the external pressure exerted by the walls and by inter-particle contacts, which together constrain the motion of the particles and permit the structure to adapt collectively to the confined environment. In figure~\ref{fig5} the black dots appearing at the particle overlaps denote the deformations associated with elastic contacts. These points correspond to the boundaries of the Voronoi polygons previously shown in figure~\ref{fig1} and are consistent with the observed adaptation phenomenon.

The percolation effects of both networks play an important role in supporting and propagating mechanical stress throughout the structure. The entropic network defines the connectivity of accessible regions of free volume, while the energetic network transmits elastic forces once contacts are established. The emergence of a percolating energetic network therefore marks the onset of mechanical cooperativity in the system.

Relaxation processes may also occur within the formed structures. These relaxation events correspond to situations in which a particle moves between adjacent regions of free volume, effectively ``jumping'' from one side of the available space to the other. This motion is driven by the compression of a particle between two others within the energetic network, while the particle remains constrained by excluded-volume overlaps within the entropic network.

This situation represents the most elementary manifestation of cooperativity in the present system: the motion of a single particle cannot occur independently but depends on the collective constraints imposed by the surrounding particles and the confining boundary. As already illustrated in figure \ref{fig4}, the competition between elastic compression and geometric constraints generates the energy barriers responsible for these cooperative rearrangements.

These results suggest several directions for further investigation, including studies of quasi-one-dimensional confined systems \cite{8,9} and the relationship between the present model and the concept of mechanical rigidity in particulate systems. In particular, the interplay between entropic constraints and elastic contacts may be related to the appearance of soft "floppy" modes recently observed in colloidal networks \cite{11}.

\section{Conclusions}

From the perspective of rigidity and force-network theories, the behavior observed here can be understood as the interplay between two distinct but coupled networks. The entropic network, formed by excluded-volume overlaps, acts as a constraint network that limits the accessible degrees of freedom without transmitting mechanical stress. As the particle growth proceeds under confinement, elastic contacts progressively activate an energetic network capable of supporting forces and storing mechanical energy.

The emergence of cooperativity is associated with the percolation of this energetic network within the pre-existing entropic constraint network. The competition between these two networks produces metastable states, mechanical adaptability, and collective response during the growth process. In this framework, the system naturally interpolates between floppy configurations dominated by geometric constraints and mechanically rigid structures in which elastic contacts transmit the stress.

In this work we introduced a minimal elastic model that captures the essential mechanisms governing the confined growth of hydrogel particles. 
The model describes particle-particle and particle-wall interactions using a purely repulsive elastic potential acting only under compression. The formulation is intentionally kept general by expressing the elastic interaction in nondimensional form with the elastic constant set to $k = 1$, which permits the model to focus on the geometric and mechanical aspects of the problem independently of the specific material parameters of a given hydrogel system.

Despite its simplicity, the model successfully reproduces the formation of structures involving different numbers of particles and explains how cooperative behavior can emerge even in the simplest case of three particles. In this configuration, the competition between excluded-volume constraints and elastic contacts generates an energy barrier in the potential landscape, producing metastable states that represent the most elementary manifestation of cooperativity in the system.

These results also provide a useful experimental-theoretical framework for exploring fundamental concepts in soft condensed matter. The growth of hydrogel particles in confined geometries offers a simple and accessible way to visualize self-organization, adaptability, and cooperativity. Self-organization appears through the spontaneous search for mechanical equilibrium, adaptability is reflected in the elastic deformation of particles under confinement, and cooperativity arises from the collective constraints imposed by the particle contacts and excluded volumes.

Finally, the model suggests several directions for further investigation. In particular, the framework may be extended to quasi-one-dimensional confined systems and to situations in which thermal fluctuations or Brownian motion are included. In this broader context, the interplay between entropic constraints and energetic contacts may provide a further insight into the mechanics of soft particulate systems and their connection to rigidity transitions, energy landscapes, and collective phenomena in confined particulate matter.

\section{Acknowledgements}
A.H. thanks 
SECIHTI for their support for the sabbatical research stay at the Department of Chemical Physics at the Institute of Physics of UNAM. L.A.P. thanks the UNAM-PAPIIT project IN111526 for its support.

\ukrainianpart
\title{Еластична модель просторово обмежених гідрогелевих частинок із конкуруючими ентропійною та енергетичною взаємодіями}

\author{А. Уерта\refaddr{label1}, Л. А. Перес\refaddr{label2}, А. Трохимчук\refaddr{label3}
}
\addresses{
	\addr{label1} Факультет фізики, Університет Веракруса, кампус Арко-Сур, Paseo 112, C.P. 91097 Халапа, Мексика
	\addr{label2} Інститут фізики, Національний автономний університет Мексики (UNAM), C.P. 04510 Мехіко, Мексика
	\addr{label3} Інститут фізики конденсованих систем НАН України імені І.Р.~Юхновського, вул. Свєнціцького 1, 79011 Львів, Україна
}

\makeukrtitle
\begin{abstract}
	У цій роботі представлено еластичну модель для дослідження ролі ентропійної та еластичної складових енергії системи в просторово обмежених гідрогелевих частинках. Розглядається квазідвовимірна система, що складається зі сферичних гідрогелевих кульок, обмежених стінками круглого контейнера, де збільшення розмірів частинок відбувається внаслідок гідратації. На основі експериментальних спостережень означено еластичний потенціал для моделювання взаємодій між частинками, а також між частинками та обмежувальною стінкою. Комп'ютерне моделювання, засноване на мінімізації енергії системи, дає змогу визначити конфігурації з найнижчою енергією, які система приймає в процесі росту розмірів частинок. Аналіз отриманих енергетичних ландшафтів демонструє наявність самоорганізації, адаптивності та кооперативності у системі, що є наслідком конкуренції між ентропійною та еластичною складовими.  
	\keywords м'яка речовина, гідрогелі, еластична модель, самоорганізація, кооперативність, ентропійна та енергетична мережі
\end{abstract}
\lastpage
\end{document}